\documentclass[%
aps, prd, reprint, show pacs, preprintnumbers, ams math, amssymb, superscriptaddress]{revtex4-1}

\usepackage{amssymb,amsmath,amstext,nicefrac}
\usepackage{graphicx}
\usepackage{dcolumn}
\usepackage{bm}
\usepackage{xspace}
\usepackage{xcolor}
\usepackage[tight]{subfigure}
\usepackage{longtable}
\usepackage{lineno}
\usepackage{pstricks}
\usepackage{appendix}
\usepackage{units}
\usepackage{rotating}

\newcommand{\dedx}{\ensuremath{\langle dE/dx\rangle}\xspace}
\def\pzpt{$(p_\mathrm{z},p_\mathrm{T})$\xspace}

\makeatother

\begin{document}

\preprint{FERMILAB-PUB-14-110-E}

\title{Measurement of Charged Pion Production Yields off the NuMI Target}

\newcommand{\FNAL}{Fermi National Accelerator Laboratory, Batavia, Illinois 60510, USA}
\newcommand{\ANL}{Argonne National Laboratory, Argonne, Illinois 60439, USA}
\newcommand{\Indiana}{Indiana University, Bloomington, Indiana 47403, USA}
\newcommand{\Harvard}{Department of Physics, Harvard University, Cambridge, Massachusetts 02138, USA}
\newcommand{\LLNL}{Lawrence Livermore National Laboratory, Livermore, California 94550, USA}
\newcommand{\USC}{Department of Physics and Astronomy, University of South Carolina, Columbia, South Carolina 29208, USA}
\newcommand{\Iowa}{University of Iowa, Iowa City, Iowa 52242, USA}
\newcommand{\UVa}{University of Virginia, Charlotteville, Virginia 22904-4714, USA}
\newcommand{\ELM}{Elmhurst College, Elmhurst, Ilinois 60126, USA}
\newcommand{\IIT}{Illinois Institute of Technology, Chicago, Illinois 60616 USA}
\newcommand{\Purdue}{Purdue University, Lafeyette, Indiana 47907, USA}
\newcommand{\UCB}{University of Colorado, Boulder, Colorado 80309, USA}
\newcommand{\UM}{University of Michigan, Ann Arbor, Michigan 48109, USA}
\newcommand{\MI}{Muons, Inc., Batavia, Illinois 60510, USA}
\newcommand{\Wichita}{Wichita State University, Wichita, Kansas 67260, USA}
\newcommand{\Delhi}{University of Delhi, Delhi 110007, India}
\newcommand{\Panjab}{Panjab University, Chandigarh 160014, India}
\newcommand{\deceased}{Deceased.}
\newcommand{\COE}{COE College, Cedar Rapids, Iowa 52402, USA}
\newcommand{\Sutcu}{Now at Kahramanmaras Sutcu Imam University, Kahramanmaras, Turkey}
\newcommand{\CalPoly}{Now at California Polytechnical State University}
\newcommand{\Queens}{Now at Queen's University, Ontario, Canada}
\newcommand{\Mustafa}{Now at Mustafa Kemal University, Hatay, Turkey}
\newcommand{\PNNL}{Now at Pacific Northwest National Laboratory, Richland, Washington, USA}
\newcommand{\IITNow}{Now at Illinois Institute of Technology, Chicago, Illinois, USA}
\newcommand{\Pitt}{Now at University of Pittsburgh, Pittsburgh, Pennsylvania, USA}
\newcommand{\BOG}{Also at Bogazici University, Istanbul, Turkey}
\newcommand{\FNALNow}{Now at Fermi National Accelerator Laboratory, Batavia, Illinois, USA}

\affiliation{\ANL}
\affiliation{\COE}
\affiliation{\UCB}
\affiliation{\Delhi}
\affiliation{\ELM}
\affiliation{\FNAL}
\affiliation{\Harvard}
\affiliation{\IIT}
\affiliation{\Indiana}
\affiliation{\Iowa}
\affiliation{\LLNL}
\affiliation{\UM}
\affiliation{\MI}
\affiliation{\Panjab}
\affiliation{\Purdue}
\affiliation{\USC}
\affiliation{\UVa}
\affiliation{\Wichita}

\author{J.~M.~Paley}
\affiliation{\ANL}

\author{M.~D.~Messier}
\affiliation{\Indiana}

\author{R.~Raja}
\thanks{\deceased}
\affiliation{\FNAL}

\author{U.~Akgun}
\affiliation{\Iowa}
\affiliation{\COE}

\author{D.~M.~Asner}
\thanks{\PNNL}
\affiliation{\LLNL}

\author{G.~Aydin}
\thanks{\Mustafa}
\affiliation{\Iowa}

\author{W.~Baker}
\affiliation{\FNAL}

\author{P.~D.~Barnes, Jr.~}
\affiliation{\LLNL}

\author{T.~Bergfeld}
\affiliation{\USC}

\author{L.~Beverly}
\affiliation{\FNAL}

\author{V.~Bhatnagar}
\affiliation{\Panjab}

\author{B.~Choudhary}
\affiliation{\Delhi}

\author{E.~C.~Dukes}
\affiliation{\UVa}

\author{F.~Duru}
\affiliation{\Iowa}

\author{G.~J.~Feldman}
\affiliation{\Harvard}

\author{A.~Godley}
\affiliation{\USC}

\author{N.~Graf}
\thanks{\Pitt}
\affiliation{\Indiana}

\author{J.~Gronberg}
\affiliation{\LLNL}

\author{E.~G\"{u}lmez}
\thanks{\BOG}
\affiliation{\Iowa}

\author{Y.~O.~G\"{u}naydin}
\thanks{\Sutcu}
\affiliation{\Iowa}

\author{H.~R.~Gustafson}
\affiliation{\UM}

\author{E.~P.~Hartouni}
\affiliation{\LLNL}

\author{P.~Hanlet}
\affiliation{\IIT}

\author{M.~Heffner}
\affiliation{\LLNL}

\author{D.~M.~Kaplan}
\affiliation{\IIT}

\author{O.~Kamaev}
\thanks{\Queens}
\affiliation{\IIT}

\author{J.~Klay}
\thanks{\CalPoly}
\affiliation{\LLNL}

\author{A.~Kumar}
\affiliation{\Panjab}

\author{D.~J.~Lange}
\affiliation{\LLNL}

\author{A.~Lebedev}
\affiliation{\Harvard}

\author{J.~Ling}
\affiliation{\USC}

\author{M.~J.~Longo}
\affiliation{\UM}

\author{L.~C.~Lu}
\affiliation{\UVa}

\author{C.~Materniak}
\affiliation{\UVa}

\author{S.~Mahajan}
\affiliation{\Panjab}

\author{H.~Meyer}
\affiliation{\Wichita}

\author{D.~E.~Miller}
\affiliation{\Purdue}

\author{S.~R.~Mishra}
\affiliation{\USC}

\author{K.~Nelson}
\affiliation{\UVa}

\author{T.~Nigmanov}
\thanks{\Pitt}
\affiliation{\UM}

\author{A.~Norman}
\affiliation{\FNAL}
\affiliation{\UVa}

\author{Y.~Onel}
\affiliation{\Iowa}

\author{A.~Penzo}
\affiliation{\Iowa}

\author{R.~J.~Peterson}
\affiliation{\UCB}

\author{D.~Rajaram}
\affiliation{\IIT}
\affiliation{\UM}

\author{D.~Ratnikov}
\affiliation{\IIT}

\author{C.~Rosenfeld}
\affiliation{\USC}

\author{H.~Rubin}
\affiliation{\IIT}

\author{S.~Seun}
\affiliation{\Harvard}

\author{A.~Singh}
\affiliation{\Panjab}

\author{N.~Solomey}
\affiliation{\Wichita}

\author{R.~A.~Soltz}
\affiliation{\LLNL}

\author{Y.~Torun}
\affiliation{\IIT}

\author{K.~Wilson}
\affiliation{\USC}

\author{D.~M.~Wright}
\affiliation{\LLNL}

\author{Q.~K.~Wu}
\affiliation{\USC}

\collaboration{The MIPP Collaboration}
\noaffiliation

\date{\today}

\begin{abstract}
The fixed-target MIPP experiment, Fermilab E907, was designed to measure the production of hadrons from the collisions of hadrons of momenta ranging from 5 to 120 GeV/c on a variety of nuclei.  These data will generally improve the simulation of particle detectors and predictions of particle beam  fluxes at accelerators.  The spectrometer momentum resolution is between 3 and 4\%, and particle identification is performed for particles 
ranging between 0.3 and 80 GeV/c using $dE/dx$, time-of-flight and Cherenkov radiation measurements.  MIPP 
collected $1.42 \times10^6$ events of 120 GeV Main Injector protons striking a target used in the NuMI facility at Fermilab.  The data have been analyzed and we present here charged pion yields per proton-on-target determined in bins of longitudinal and transverse momentum between 0.5 and 80 GeV/c, with combined statistical and systematic relative uncertainties between 5 and 10\%.

\end{abstract}

\maketitle

\begin{figure*}[htpb]
   \centering
   \includegraphics[width=0.9\textwidth]{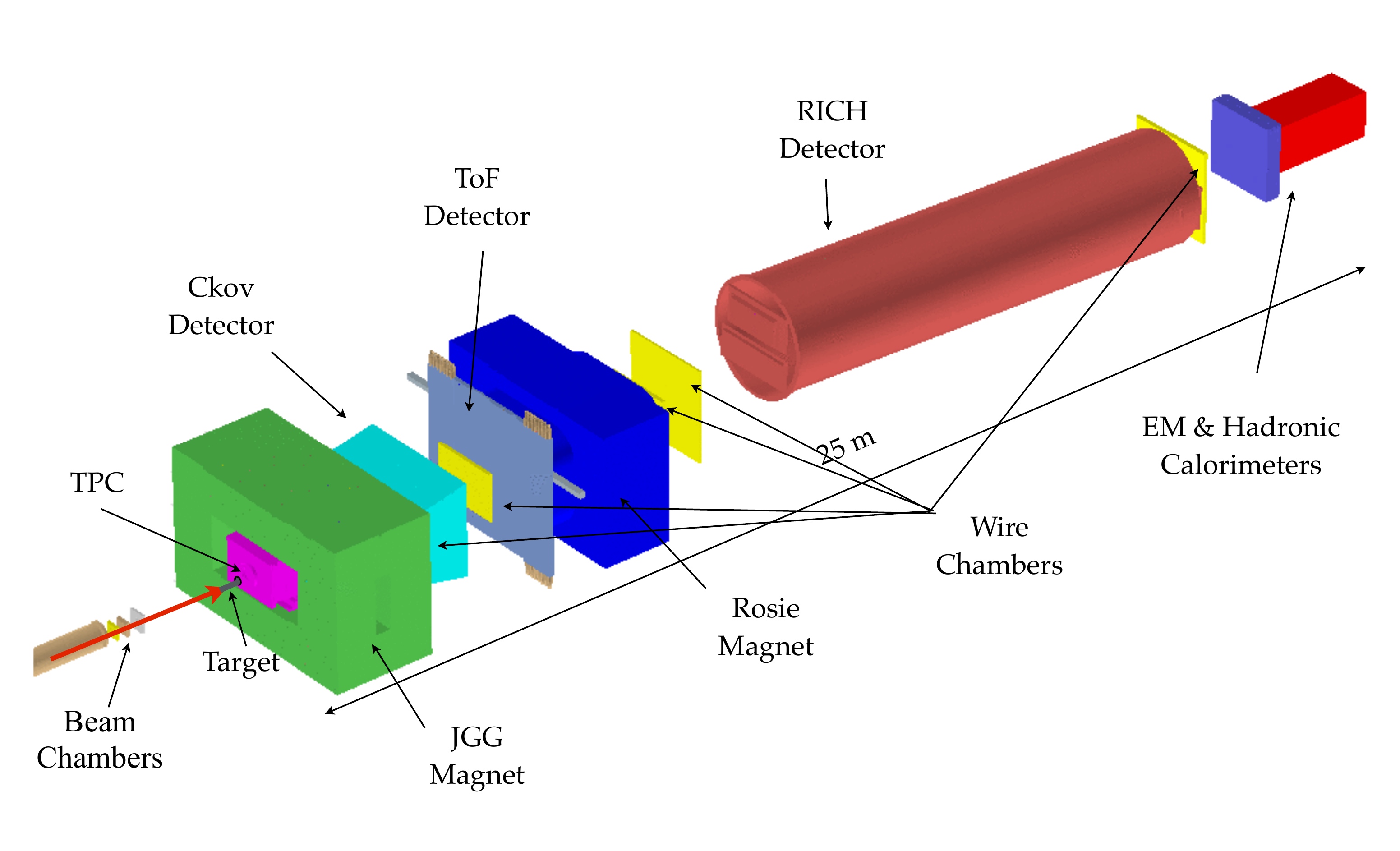}
   \caption{Schematic view of the MIPP spectrometer.}
   \label{fig:MIPPSpect}
\end{figure*}

\section{Introduction}
A growing number of neutrino experiments conducted at proton accelerators derive their neutrino beams from 
horn-focused beams of pions and kaons which result from proton-nucleus collisions in low-Z materials.  At the 
Neutrinos at the Main Injector (NuMI) facility at Fermilab, hadron production uncertainties in Monte Carlo (MC) simulations generally dominate 
the uncertainties of the neutrino flux predictions at the level of 15--20\%, and are a limiting factor in the neutrino 
and anti-neutrino cross-section measurements being done by many NuMI-based experiments
~\cite{ref:MINOS,ref:MINERvA1,ref:MINERvA2}.  
One of the goals of the Main Injector Particle Production (MIPP) experiment 
was to measure the hadron production yield off of an actual NuMI target with 120 GeV/c protons from 
the Main Injector (MI) to within 5\% in order to verify and validate the Monte Carlo calculations of the NuMI flux.  Here we report a measurement of charged pion yield per 120 GeV/c proton-on-target in approximately 120 bins of longitudinal and transverse momentum.  The statistical and systematic uncertainties in most bins are between 5 and 10\%.

\section{The MIPP Spectrometer}

For the MIPP experiment, an open geometry 25~m long spectrometer was assembled with two dipole magnets for momentum determination, a 1.5 m long time-projection chamber~\cite{ref:TPC} (TPC) located just downstream of the interaction region, and 3 drift chambers (DCs) and 2 multiwire proportional chambers (MWPCs) located further downstream for particle tracking.  The TPC sits inside the most upstream dipole magnet, and targets are placed just upstream of the TPC.  Three wire chambers~\cite{ref:BeamChambers} (BCs) positioned across 36 meters upstream of the target are used to track incident beam particles.  A schematic of the spectrometer is shown in Fig.~\ref{fig:MIPPSpect}.  

The electric field used to drift ionization electrons produced in the MIPP TPC is aligned the 
MIPP was designed to provide particle identification (PID) with $2-3\sigma$ separation across the momentum range of a few hundred MeV/c to greater than 80 GeV/c using \dedx information from the TPC (0.2--1.2 GeV/c), a plastic scintillator-based time-of-flight (ToF) detector (0.5--2.5 GeV/c), a segmented gas Cherenkov detector~\cite{ref:Ckov} (2--20 GeV/c) and a gas ring imaging Cherenkov~\cite{ref:RICH} (RICH) detector (4--80 GeV/c).  Electromagnetic and hadronic calorimeters are used to measure forward-going neutrons and photons~\cite{ref:MIPPNeutron}.  The high multiplicities present in this data set complicate the use of the Cherenkov and ToF detectors, and this analysis relies on measurements from the TPC and RICH detectors.  Measurements from the ToF detector are used to estimate backgrounds.

\section{Target and Incident Beam}
The NuMI target used in this measurement was a spare target that was eventually used by the NuMI complex after the MIPP data run.  The target was designed for operation in the low-energy configuration of the NuMI beamline and consists of a 90-cm long, 3-cm diameter aluminum vacuum can encompassing 47 2-cm thick graphite slabs, adding up to two nuclear interaction lengths of material.  The downstream end of the tube was positioned 8 cm away from the upstream end of the TPC active volume.  

The incident beam was 120 GeV/c protons, slow-extracted directly from the MI.  A pinhole collimator was used to reduce the incident MI proton beam flux by 8 orders of magnitude, such that the rate of incident beam pileup in the target over the 16 $\mu s$ required to read out the TPC was reduced to a few percent.  

\begin{figure}[t]
   \centering
   \includegraphics[width=0.45\textwidth]{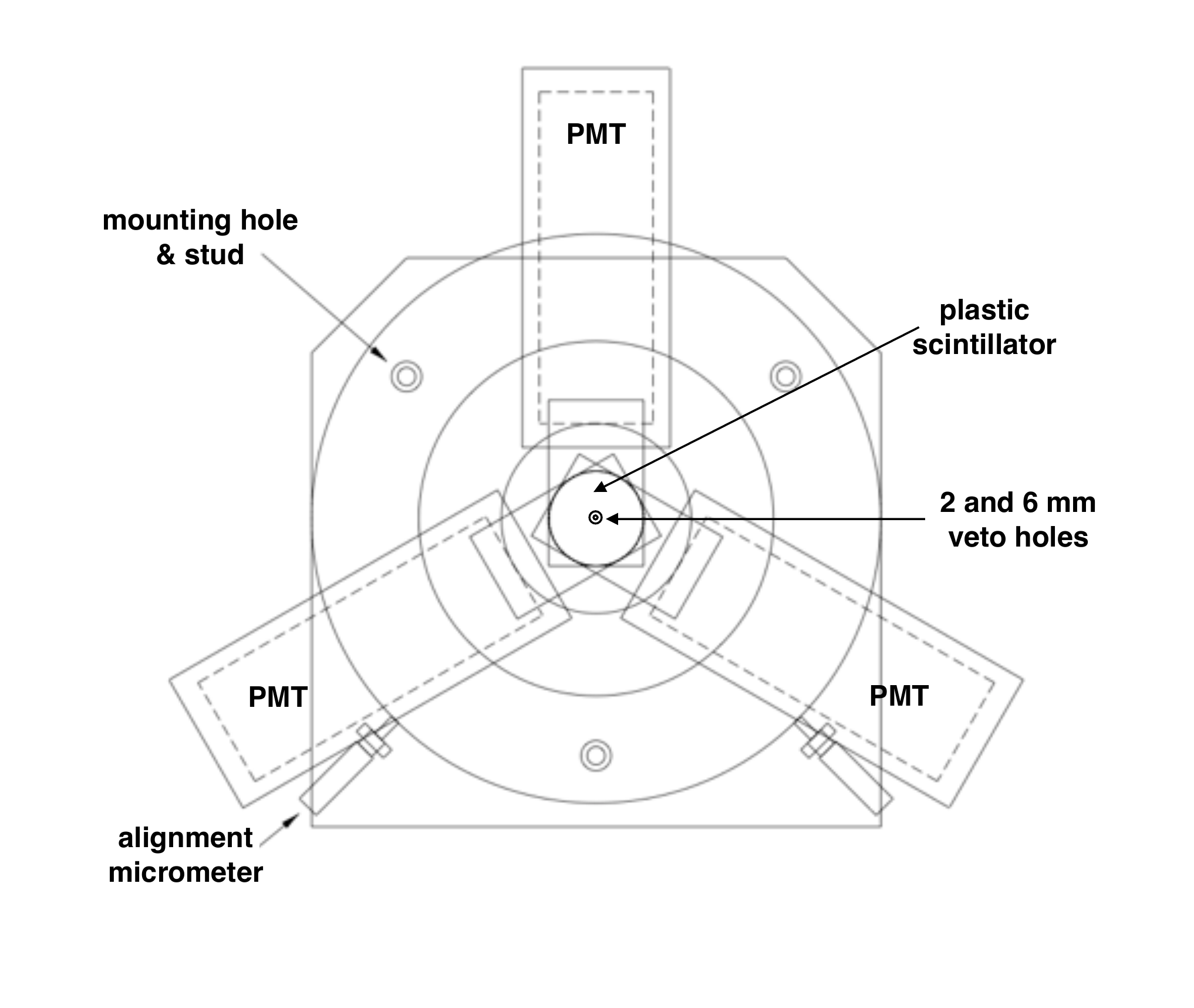}
   \includegraphics[width=0.45\textwidth]{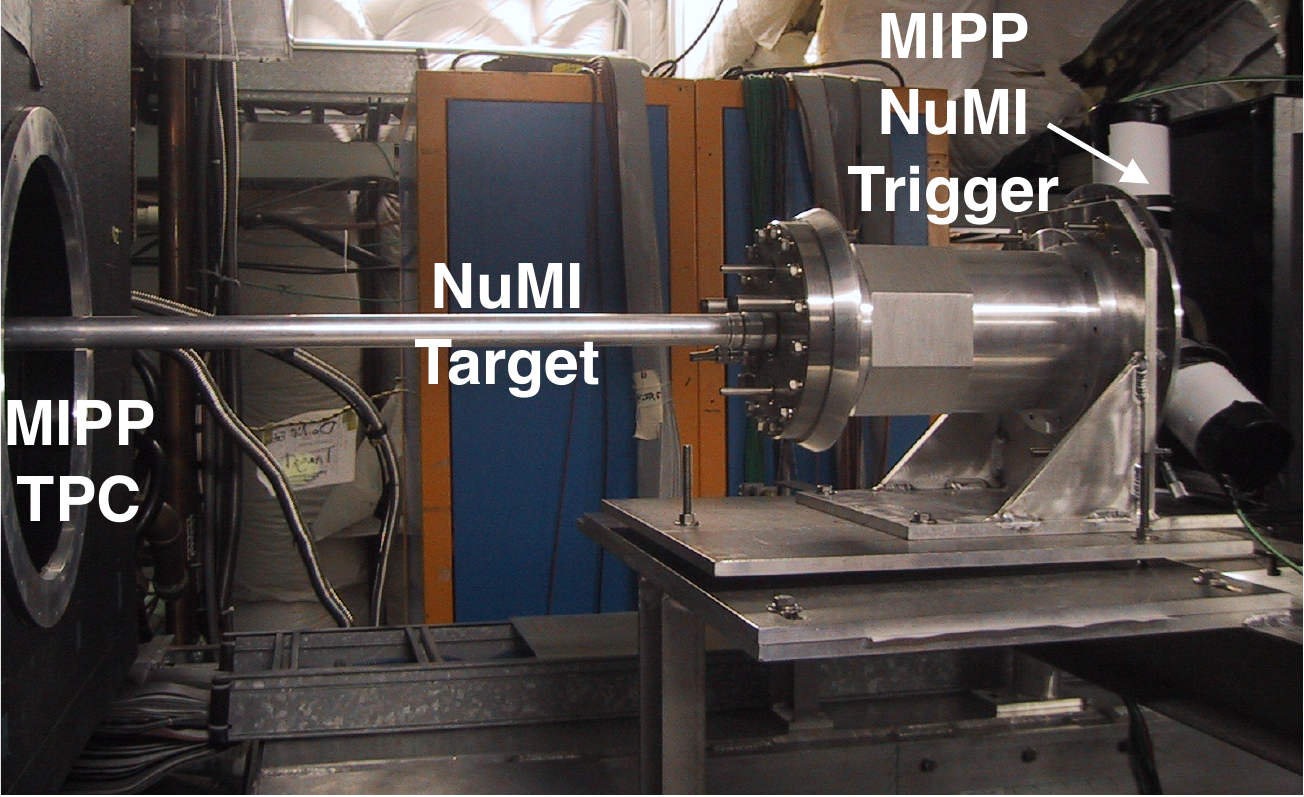}
   \caption{Top: schematic of the MIPP NuMI trigger system.  Bottom: photo of the NuMI trigger mounted in the MIPP experiment, with the trigger system mounted on the upstream face of the target.}
   \label{fig:numitrigger}
\end{figure}

\begin{figure}[t]
   \centering
   \includegraphics[width=0.45\textwidth]{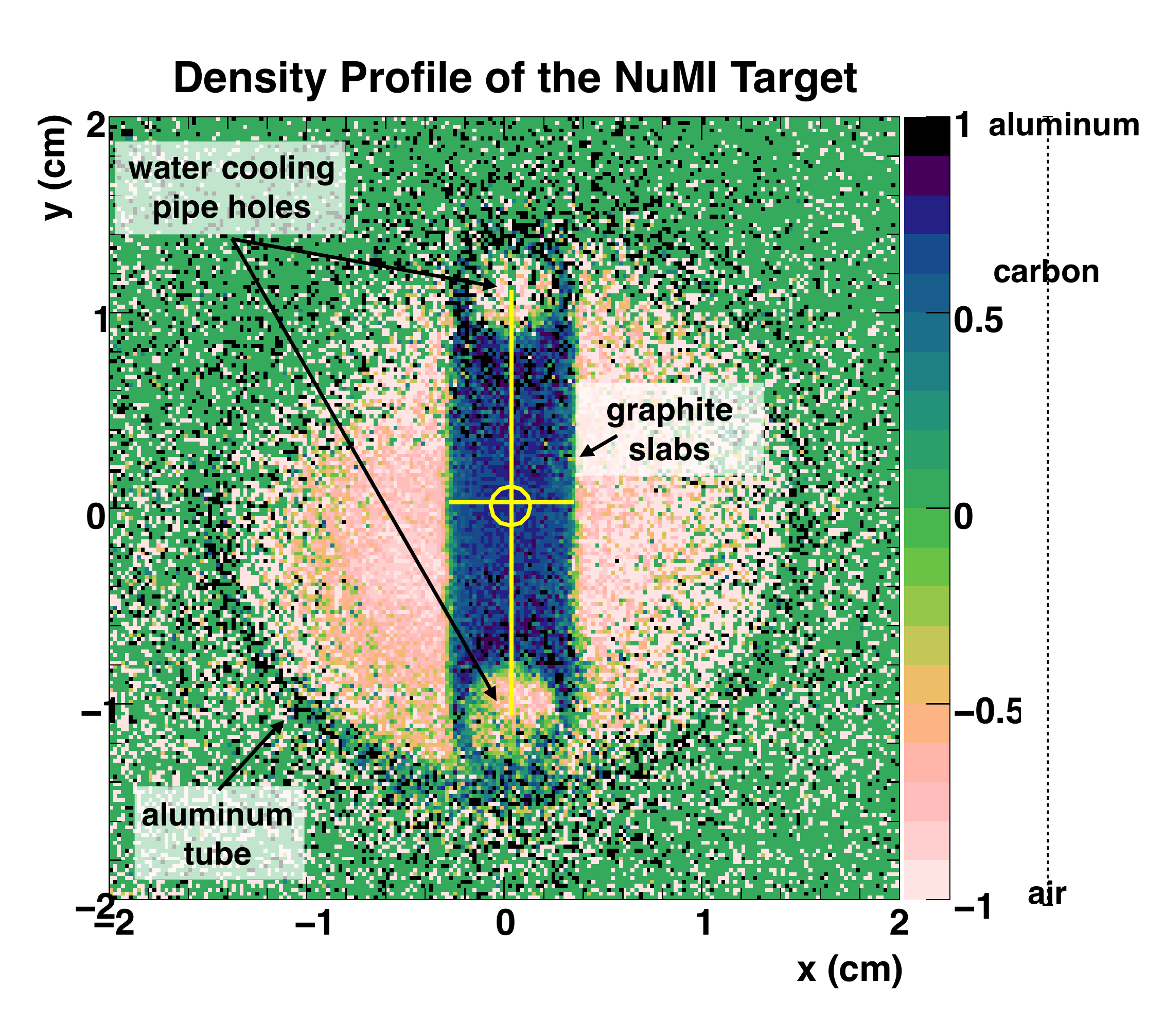}
   \caption{Measured probability that an incoming beam proton interacts with the target material as a function of the position of the proton at the upstream face of the target.  Cross-hairs represent the center of the graphite slabs, and the circle represents the position of the 2-mm trigger hole at the face of the target.}
   \label{fig:trgtprof}
\end{figure}

In order to ensure that the incident beam was centered on the face of the target, a NuMI trigger detector consisting of three thin ($\lambda_L < 0.5\%$) overlapping pieces of plastic scintillator was 
mounted on the upstream face of the NuMI target.  The light from each of the scintillator pieces was detected by a photomultiplier tube (PMT).  
The middle and most downstream pieces of scintillator had circular holes 2 mm and 6 mm in diameter, respectively, drilled in the 
center.  A ``2-mm'' trigger was formed via a coincidence of a signal from the upstream beam counters with the signal from the upstream scintillator of the NuMI trigger and the absence of signals from the middle and downstream scintillators of the NuMI trigger.  In a similar fashion, a second ``6-mm'' trigger was defined by a signal from the upstream and middle scintillators and the absence of a signal from the downstream scintillator.  This second trigger was pre-scaled such that the profile of the beam of protons striking the NuMI target in the MIPP experiment matched the Gaussian profile of the 1 mm beam width observed in the NuMI beam facility.  Incident beam pileup was reduced via a veto if two triggers fire within 1 $\mu$s of each other.

Shown in Fig.~\ref{fig:trgtprof} is the measured probability that an incoming beam proton interacts with the target material as a function of the position of the proton at the upstream face of the target.  The data used to generate Fig.~\ref{fig:trgtprof} were collected with a minimum bias trigger during the NuMI target commissioning period and are not used in this analysis.  During 
this commissioning period the beam was randomly swept across the upstream face of the target and some regions of the target were exposed to more beam than others.  As a result, the outlines of the aluminum vacuum tube, water cooling pipe holes and graphite slabs is non-uniform.  The reconstructed positions of the incident beam particles on the upstream face of the target are measured using data from the BCs.  The reconstruction of the incident beam trajectory uses the charge and time information recorded in the 4 1-mm pitch wire planes of each BC, resulting in 0.1 mm track position reconstruction resolution at the upstream face of the target.  The data recorded during the target scan involved random beam  positioning, and some regions of the upstream face of the target received different levels of beam exposure.

The color/shade in Fig.~\ref{fig:trgtprof} is determined from the normalized difference between the number of events with greater than 3 secondary tracks observed in the downstream spectrometer and the number of events with a single beam proton in the downstream spectrometer.  Darker regions correspond to higher density materials, lighter regions to less dense materials.  The graphite slabs located inside the NuMI target are clearly 
visible, with holes at the top and bottom where aluminum water cooling pipes, empty and exposed to air during data taking in the MIPP experiment, run along the length of the target.  The outer aluminum tube containing the 
graphite and water cooling pipes is also visible.  The graphite slabs were 
actually found to be rotated $3^{\circ}$ about the longitudinal axis; this rotation
has been removed in Fig.~\ref{fig:trgtprof}.  

The cross-hairs in Fig.~\ref{fig:trgtprof} represent the width and height
of the graphite slabs.  The positions of the cross-hairs were determined 
by fitting the edges of the $x$- or $y$-projection of the plot.  The measured
width of the graphite slabs is 6.36 mm; the technical specification of the width of the slabs in the NuMI
target is 6.4 mm.  

The center of the 2-mm wide circle near the center of Fig.~\ref{fig:trgtprof} represents the mean distribution of reconstructed incident beam positions on the face of the target for 2-mm triggered events.  The beam center position is offset by 0.045 mm in the horizontal direction, and 0.174 mm in the vertical direction, from the center of the target.  

\section{Simulation}
MC simulations are used to determine event and track selection criteria and acceptance, track reconstruction and particle identification efficiency corrections in the analysis.  Interactions of 120 GeV/c protons striking the NuMI target in the MIPP experiment are simulated using Fluka (v2005)~\cite{ref:Fluka} for event generation (e.g., 120 GeV/c proton interactions on the NuMI target) and GEANT3~\cite{ref:GEANT3} for particle trajectory tracking.  
The Fluka simulation generates primary, secondary, tertiary, etc. interactions of particles within the target and housing, 
and has a detailed geometric description of the NuMI target, the same geometry employed by the MINOS 
experiment.  The simulated beam profile at the target has a full-width at half-maximum (FWHM) of 0.26 mm in the horizontal direction and 0.29 mm in the vertical direction centered on the upstream face of the target, which are the measured widths of the beam spot size on the NuMI target in the NuMI facility, and is within 10\% of of the measured beam spot size on the NuMI target in the MIPP experiment.  Fluka tracks each particle produced in the target until it reaches the surface of the target, which is 
the outer edge of the aluminum pipe encasing the graphite slabs of the target.  The next stage of the MC 
generation is GEANT3, which uses the output of the Fluka simulation as input, and tracks each particle taking 
into account multiple scattering, energy loss and decays through a detailed geometric description of the MIPP 
spectrometer.  GEANT ``hits'' are recorded in each detector volume until the particle either loses all energy or is 
well outside the volume of the MIPP spectrometer.  The last stage of the MC simulation converts the GEANT hits 
to simulated digital signals, tuned to match data recorded in the experiment.

\section{Particle Trajectory Reconstruction}

\begin{figure}[t]
   \centering
   \includegraphics[width=0.5\textwidth,trim= 15 15 5 15 mm]{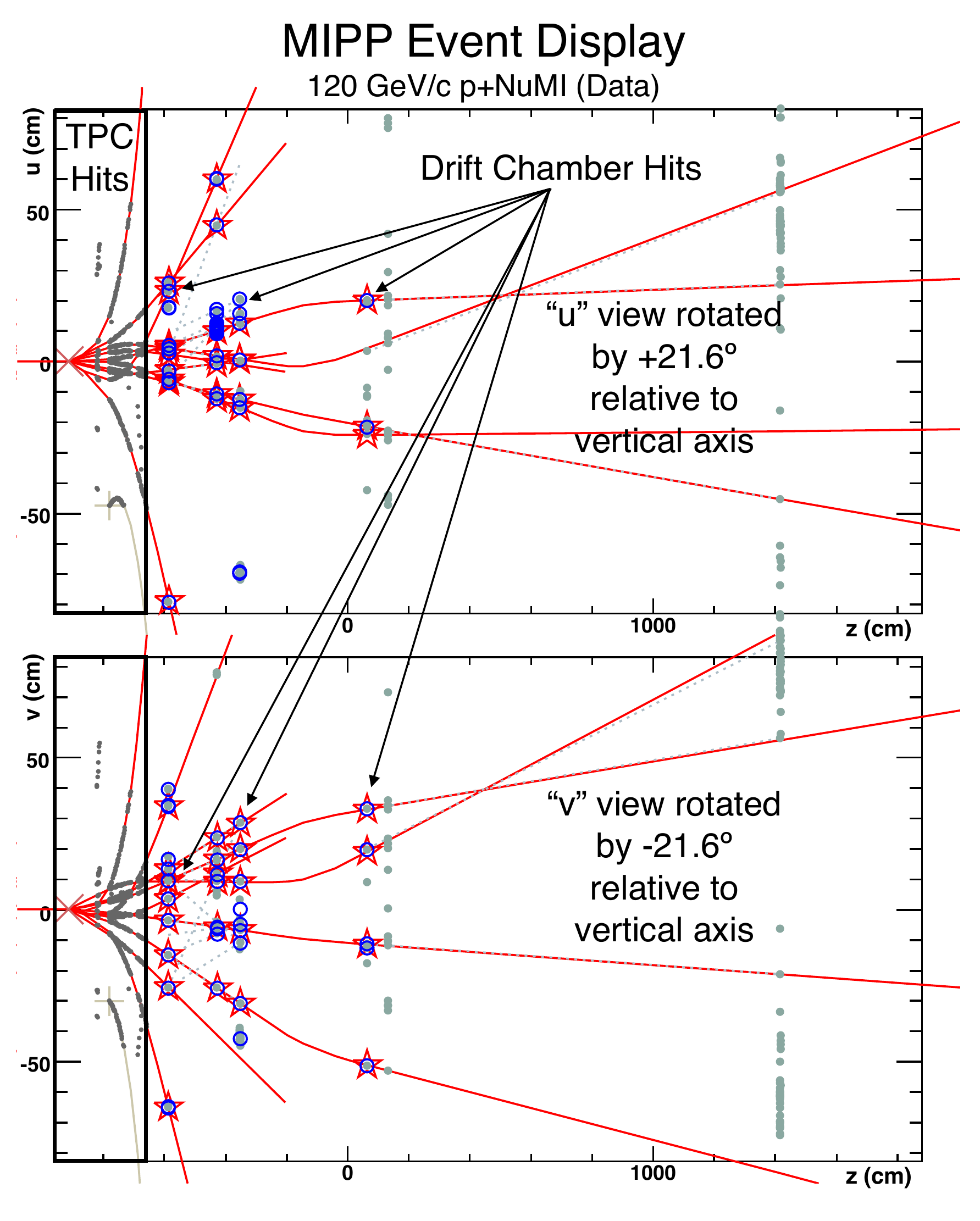}
   \caption{MIPP event display of real data showing the secondary particle reconstruction using recorded hits in the TPC and downstream drift and proportional wire chambers.  The incident proton beam enters from the left.  The gray points are the hits recorded in each detector and the solid red lines represent the reconstructed trajectories of the secondary particles emanating from the target.  The views in the top and bottom have been rotated by $\pm21.6^\circ$ to display the plane-view of hits in two of the four planes of the downstream DCs.  The blue circles represent DC hits in each view, and the red stars represent hits in each view that have been associated with a reconstructed track. }
   \label{fig:evd}
\end{figure}

The MIPP event reconstruction includes reconstruction of the trajectory of the primary beam particle using data 
from three wire chambers located upstream of the target, reconstruction of the secondary particles originating 
from the target, and matching the 
secondary particles to data recorded in specific channels in the ToF, Cherenkov, RICH, EMCal and HCal 
detectors.  The secondary particle trajectories are reconstructed by merging reconstructed track segments from 
hits in the TPC detector with track segments formed 
from hits in the downstream DCs and PWCs.  Fig.~\ref{fig:evd} shows an event display of NuMI target data recorded in the MIPP experiment.
In the analysis of the NuMI target data, Monte Carlo simulation studies indicate that the momentum resolution is $3-5\%$, and the transverse momentum resolution is less than 20 MeV/c for all momenta. 

\subsection{Particle Identification}

\begin{figure}[thp]
   \centering
   \includegraphics[width=0.5\textwidth, trim= 5 5 5 5 mm]{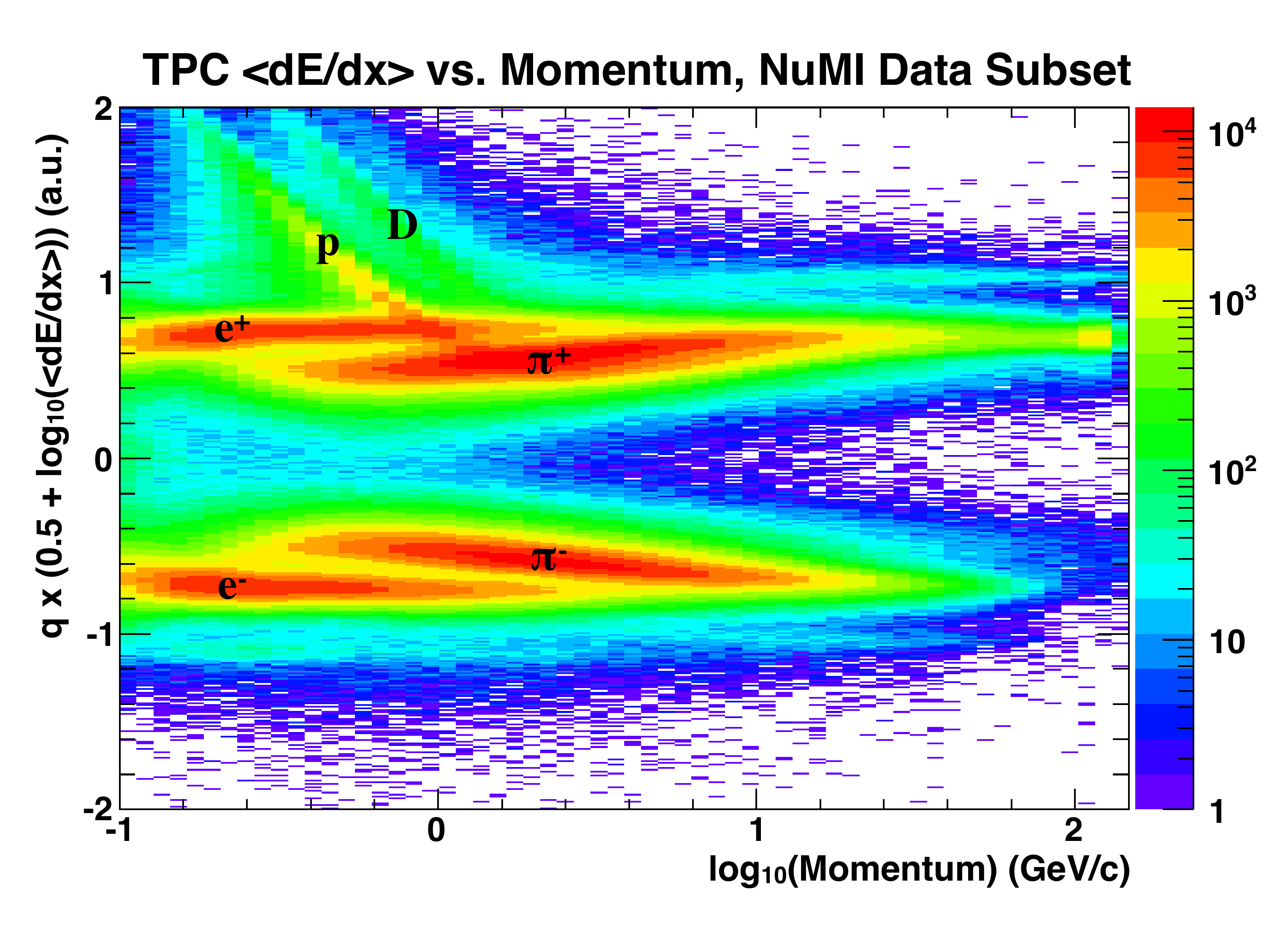}
   \caption{Distribution of measured charge $\times$ \dedx versus $\log_{10}(p)$.  Colors represent the density of particles at the reconstructed momentum and \dedx, and bands for different particle types are clearly visible.}
   \label{fig:dedxVsMom}
\end{figure}

The \dedx is determined for every reconstructed track from TPC hits based on the charge recorded on 8 mm $\times$ 12 mm charge-sensitive pads in the readout-plane of the TPC.  Time-dependent corrections, relevant on the order of hours to days, are applied to the data to account for monitored changes in water-vapor and oxygen contamination in the TPC gas.  
Tracks in this analysis had 20--90 associated TPC hits providing \dedx resolutions between 15 and 25\%.

Reconstructed tracks are matched to hits recorded in the ToF.  Temperature-dependent and cross-talk corrections are applied to the ToF data, and improve the timing resolution of the detector from 1.2 ns to 0.4 ns.  As a result, the ToF data provide $\pi-p$ separation up to about 2 GeV/c.  The recorded flight time, $\Delta t$, in the ToF detector is converted to an invariant $m^2$ via
\begin{equation}
m^2 = p^2 \left(\frac{c^2\Delta t^2}{\Delta L^2} - 1\right)
\end{equation}
where $p$ is the reconstructed momentum of the particle and $\Delta L$ is the reconstructed flight distance from the target to the ToF detector.  Finite resolutions in $\Delta t$ and $\Delta L$ occasionally result in negative values of $m^2$.  Due to the high multiplicities of secondaries in the NuMI data set, approximately 50\% of all ToF data recorded per event are a result of 2 or more particles passing through the same ToF bar.  It is therefore impossible to disentangle these particles in the ToF data.  The remaining 50\% of the data from the ToF detector are used in conjunction with the \dedx measurements as described in Section \ref{section:TPCdEdXMeas}.

\begin{figure}[!thp]
   \centering
   \includegraphics[width=0.45\textwidth]{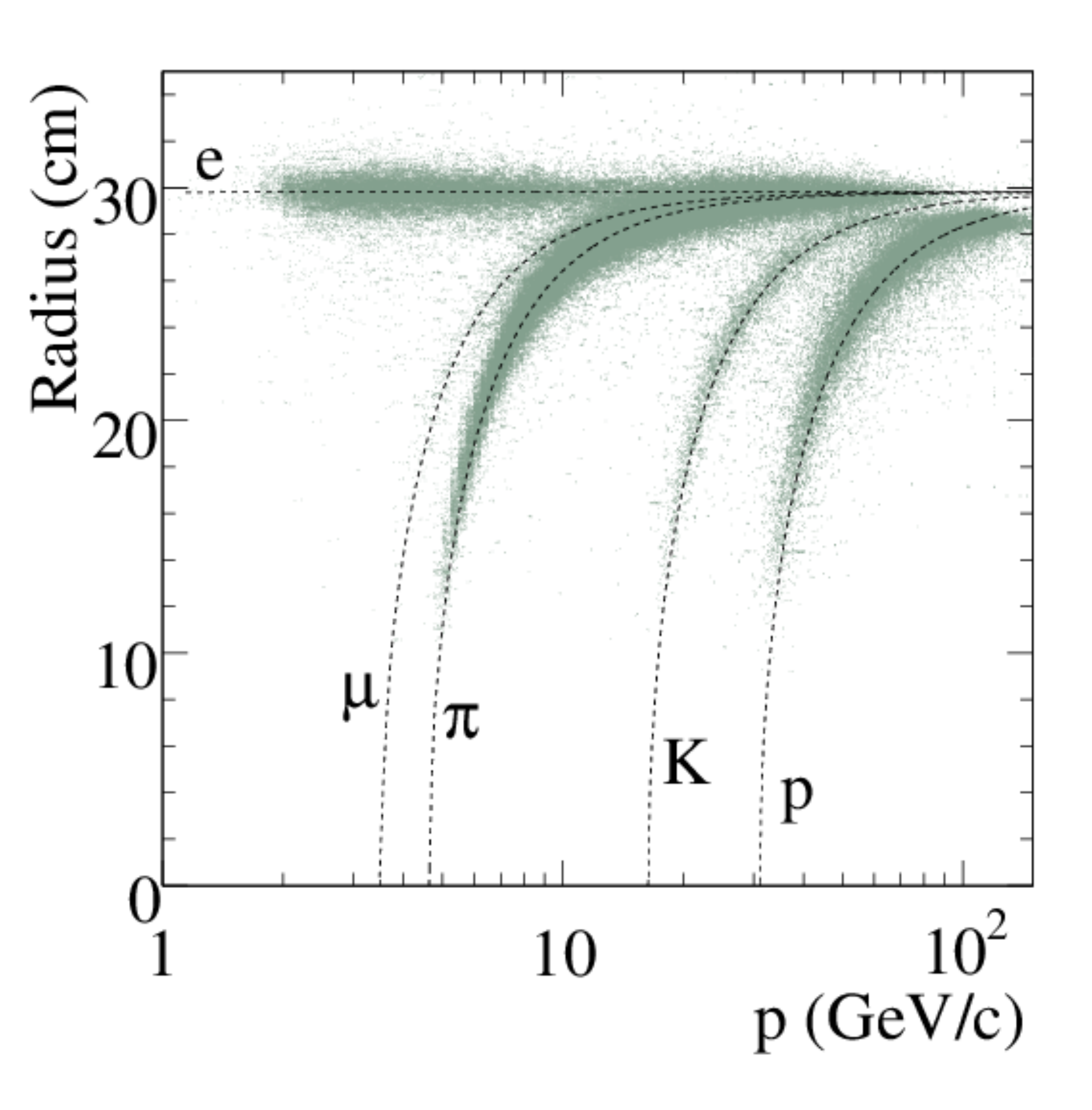}
   \caption{Reconstructed RICH ring radius versus track momentum for positively charged tracks in the NuMI target data set.  Gray points are measurements for individual tracks, and the predicted bands for the different particle types are superimposed as dashed curves.}
   \label{fig:richRingVsMom}
\end{figure}

Particles in the RICH detector produce light cones which are reflected to form a ring of light on an array of approximately 2300 1/2-inch PMTs.  The high segmentation of the RICH detector allows multiple rings to be clearly distinguished and matched to reconstructed tracks.  The efficiency of matching a reconstructed track to a reconstructed RICH ring is $>98\%$, and the high multiplicity of secondaries is not an issue for this detector.  Fig.~\ref{fig:richRingVsMom} is a scatter plot of the reconstructed RICH ring radius versus matched reconstructed particle momentum.  The predicted bands for different particle types are drawn with dashed curves.  Given a particle's momentum, $p$, the matched RICH ring radius is converted to an invariant $m^2$ assuming the small-angle approximation:
\begin{equation}
m^2 \simeq p^2 n^2\left(1 - \left(\frac{r}{L}\right)^2\right) - p^2
\end{equation}
where $n$ is the refractive index (1.00045 for the CO$_2$ used in the MIPP RICH detector), $r$ is the reconstructed RICH ring radius and $L$ is the length of the RICH radiator volume (990 cm).

\section{Analysis}

This analysis is a measurement of the pion yield off the NuMI target, $N_\pi$\pzpt per proton-on-target (POT).  POT is defined as the number of reconstructed events that pass the event selection described in Section \ref{Sec:EventSelection}, and includes events where the proton does not interact in the target.  Yields are extracted from TPC \dedx and RICH $m^2$ distributions.  Corrections are applied to each measurement to account for spectrometer geometric acceptance, track reconstruction efficiency, PID detector geometric acceptance and PID detector efficiency:

\begin{equation}
N_\pi(p_\mathrm{z},p_\mathrm{T}) = \frac{N_\pi^\mathrm{meas}(p_\mathrm{z},p_\mathrm{T})}{\epsilon^\mathrm{spect}_\mathrm{accept} \times \epsilon^\mathrm{reco}_\mathrm{eff} \times \epsilon^\mathrm{PID}_\mathrm{accept} \times \epsilon^\mathrm{PID}_\mathrm{eff}}
\end{equation}
In general, unless otherwise noted, the measurement and calculations of corrections to be applied to the data are done for positive and negative particles separately.  

Prompt muon production dominated by decays of charm mesons is known to be a $\ll 0.1\%$ effect, thus we treat all muons as arising from pion decay and pion MC PID distributions include contributions from muons.  MC studies confirm that this contribution is negligible in all \pzpt bins in the analysis.  

\subsection{Momentum Calibration}
A small ($< 1\%$) correction, based on a comparison between reconstructed and true momenta of MC tracks, is applied to 
the reconstructed momenta of tracks through the MIPP spectrometer to account for energy loss and scattering, as well as biases introduced by the reconstruction algorithm.  The overall momentum scale is calibrated using reconstructed primary beam protons that pass through the target and reconstructed K$^0_\mathrm{S}$ mass from pairs of oppositely charged tracks produced off the target.  The primary beam momentum was found to agree with the expected 119.6 GeV/c from the MI.  The invariant mass distribution of combinations of oppositely charged tracks in data showed a peak near the K$^0_\mathrm{S}$ above a flat background.  A gaussian fit to this peak had a central value $(0.85 \pm 0.08)\%$ lower than the PDG~\cite{ref:PDG} value.  The momenta of tracks that contribute to the K$^0_\mathrm{S}$ mass peak is peaked around 1 GeV/c.  A linear interpolation between these two measurements, one at 1 GeV/c (0.85\% offset) and the other at 120 GeV/c (0\% offset) is used to correct for absolute momentum.

\subsection{Event Selection}
\label{Sec:EventSelection}
Event selection in this analysis is designed to reject events with multiple incident beam particles (protons)  while requiring that the beam be centered on the NuMI target.  We require exactly one reconstructed incident beam track from data recorded in the upstream beam wire chambers, a reconstructed beam track time that falls within the expected 18.9 ns-wide window from the accelerator RF bucket, 
and a reconstructed beam track position that falls within 0.648 cm of the center of the upstream face of the target.   
Because the time to drift ionization from tracks out of the TPC volume is 16 $\mu s$ and the drift is in the vertical direction, particles traversing the center of the TPC many $\mu s$ before [after] the event trigger will have shorter [longer] recorded times and therefore appear to be well below [above] the center of the TPC.  Events with an excess of TPC tracks appearing at the top or bottom of the TPC were rejected, and MC studies, where no incident beam pileup is simulated, indicate the rejection of these events to have a negligible bias.

\subsection{Track Selection}
\label{Sec:TrackSelection}
Track selection in this analysis, applied after event selection, is designed to reject tracks that are poorly reconstructed or do not originate from the target.  We require all reconstructed tracks to be above a  ``goodness-of-fit'' (GoF) parameter threshold, derived from the trajectory fit residuals.  The GoF threshold was determined by a sharp cut-off in the GoF distribution seen in both data and MC.  All tracks must have at least 8 TPC hits, although in practice tracks with momenta in the analysis bins have $> 20$ TPC hits.  Finally, every track is required to have a trajectory that has a distance of closest approach (DoCA) to the target cylinder that is within 2 cm in $r$ and within 5 cm in $z$.  The resolution of the DoCA in $r$ is 0.7 cm and 2.8 cm in $z$.  No significant disagreement was found between the shapes of the distributions of the track selection criteria in data and MC.

\subsection{Binning}
The yields of secondary pions produced in the target are measured in bins of $p_\mathrm{z}$ and $p_\mathrm{T}$ chosen to keep the statistical uncertainty in each bin less than $5\%$ while maintaining a bin width that exceeds the momentum resolution of the spectrometer by at least a factor of 4.  A total of 150 bins are defined in this analysis; however due to limited statistics in some bins, positive [negative] pion yields are reported for 124 [119]  bins covering 0.30 to 80 GeV/c and 6 bins from 0 to 2 GeV/c in $p_\mathrm{T}$.

\subsection{Efficiency and Acceptance Corrections}
\begin{figure*}[t]
  \centering
  \subfigure[ {} Combined geometric acceptance, track reconstruction and track selection efficiency for positive pions.  ]{\label{Fig:RecoEff} 
  \includegraphics[width=0.47\textwidth]{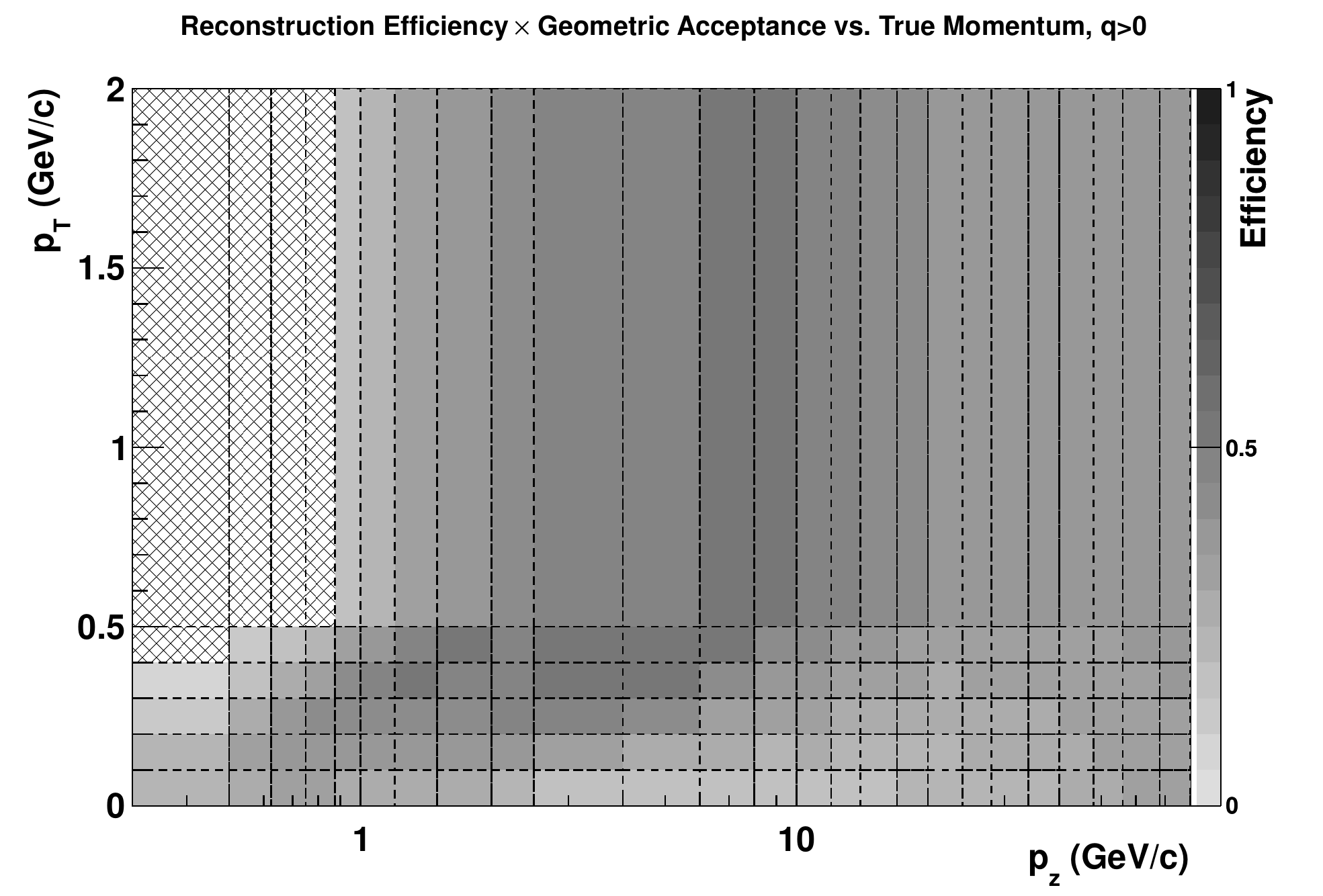}}
  \subfigure[ {} Geometric acceptance of the TPC and RICH particle identification detectors.]{\label{Fig:PIDAccept}
  \includegraphics[width=0.47\textwidth]{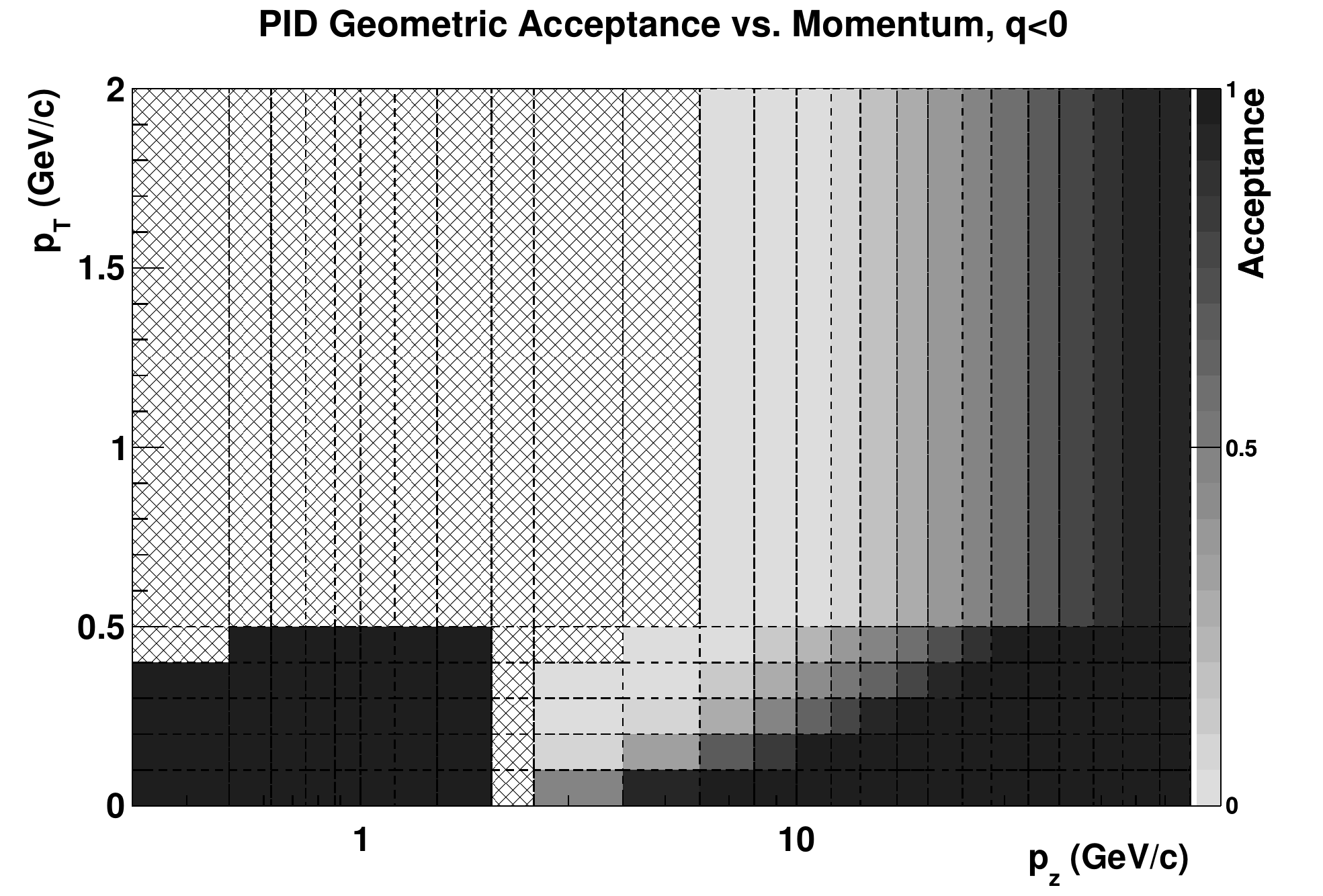}}%
  \caption{Reconstruction efficiencies and acceptances as a function of \pzpt as determined from MC simulation.  The shade represent the efficiency, darkest representing 100\% and lightest representing 0\%.  Hashed bins represent excluded measurements.}
  \label{Fig:AcceptPlots}
\end{figure*}

Geometric acceptance and track reconstruction efficiency corrections are determined using MC simulations which have detailed descriptions of the target, spectrometer and detector geometries.  The combined geometric acceptance of the spectrometer, track reconstruction efficiency and track selection efficiency is shown in Fig.~\ref{Fig:RecoEff} as a function of \pzpt.  The geometric acceptance of the PID detectors is shown in Fig.~\ref{Fig:PIDAccept}.  The shade of the boxes indicates the scale of the effect, where 100\% efficiency or acceptance is the darkest shade and the lightest shade represents 0\%.  Both plots show results for negative particles; positively charged particles have very similar efficiencies and acceptances.  All tracks are required to have a reconstructed TPC track segment.  The acceptance of the RICH requires tracks to traverse the spectrometer, which is limited to high momentum.  This is reflected in the 100\% acceptance of the TPC \dedx PID and the much smaller RICH PID acceptance.  The hashed bins have been excluded due to poor statistics.

\subsection{TPC Measurements}
\label{section:TPCdEdXMeas}
Every reconstructed track has a corresponding measurement of TPC \dedx.  In any given slice of total momentum, the 
distribution of log(\dedx) for any particle type is nearly Gaussian, and the $\log_{10}$(\dedx) distributions in narrow bins of \pzpt are 
very nearly Gaussian.  We therefore fit the $q \times$log(\dedx) distributions to a sum of six Gaussians, 2 peaks 
each for $e, \pi$ and $p$.  The resulting distribution is centered about zero by construction.  Fitting the data in this way reduces the number of free Gaussian parameters from 18 to 12:  3 means, 3 widths, and 6 amplitudes.  The fit function is rewritten as
\begin{eqnarray}
\label{Eqn:dEdx}
N(y) &  =  & A_{\pi^+}\left[f^{+}_{e\pi}G^+_e(y) + G^+_\pi(y) + f^+_{p\pi}G^+_p(y)\right]  + \nonumber \\
& & A_{\pi^-}\left[f^{-}_{e\pi}G^-_e(y) + G^-_\pi(y) + f^-_{p\pi}G^-_p(y)\right]  
\end{eqnarray}
where 
\begin{equation}
G^\pm_i = \exp\left(\frac{(y\mp y_i)^2}{2\sigma_i^2}\right)
\end{equation}
is the Gaussian function for each particle type, $y$ is the measured value of $\log_{10}$\dedx, $y_i$ is the 
mean, $\sigma_i$ is the width, $A_{\pi^\pm}$ is the fit amplitude of the pion peak, and $f_{e\pi}$ ($f_{p\pi}$) is the 
amplitude ratio of the electron (proton) peak to the pion peak.

\begin{figure}[!ht] 
   \centering
   \includegraphics[width=0.45\textwidth]{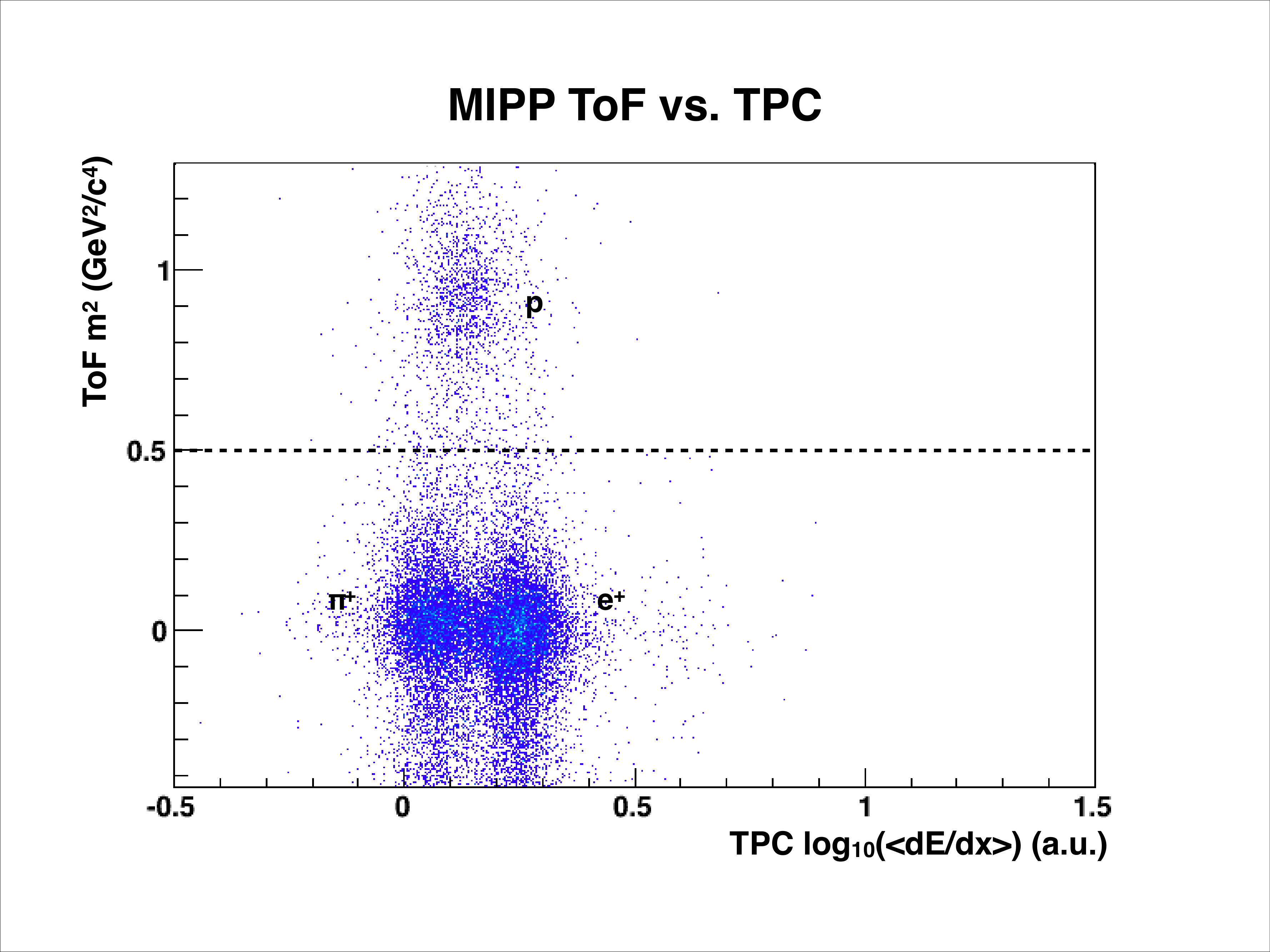} 
   \caption{ToF $m^2$ vs.~TPC \dedx for $1.2 \le p_\mathrm{z} < 1.5, 0.0 \le p_\mathrm{T} < 0.15$ GeV/c for tracks with isolated hits in the ToF.  Note that the protons are clearly distinguished from pions and electrons in the ToF, whereas they fall under the pion and electron peaks in the TPC \dedx distribution.}
   \label{fig:ToFvsTPC}
\end{figure}

One feature of the \dedx distributions is that at higher momenta very large fractions of the proton peak fall under the pion peak and the fit to 6 peaks fails.  However, in this range, the protons are clearly distinguished from pions and electrons in the ToF.  Fig.~\ref{fig:ToFvsTPC} shows the reconstructed $m_\mathrm{ToF}^2$ versus the \dedx of tracks where no other reconstructed trajectories traverse the ToF scintillator bar.  The protons are very clearly visible in the ToF, whereas these protons fall under the pion and electron \dedx peaks on the x-axis.  The $\pi/p$ fraction is determined from these data by assuming all particles with 
$m_\mathrm{ToF}^2$ between 0.5 and 1.2 GeV${^2}$/c$^{4}$ are protons, and fitting the \dedx distributions for tracks with ToF $m_\mathrm{ToF}^2$ below 0.5.  The results of these fits are then used as a constraint to fits of the TPC \dedx distribution for tracks with momentum greater than 0.88 GeV/c where ToF data may not be used because of multiple trajectories traversing the same ToF scintillator bar.

\begin{figure*}[!ht] 
   \centering
   \subfigure[{} $0.5 \le p_\mathrm{z} < 0.62, 0.10 \le p_\mathrm{T} < 0.20$ GeV/c.]{\label{fig:fitdedx7}\includegraphics[width=0.45\textwidth]{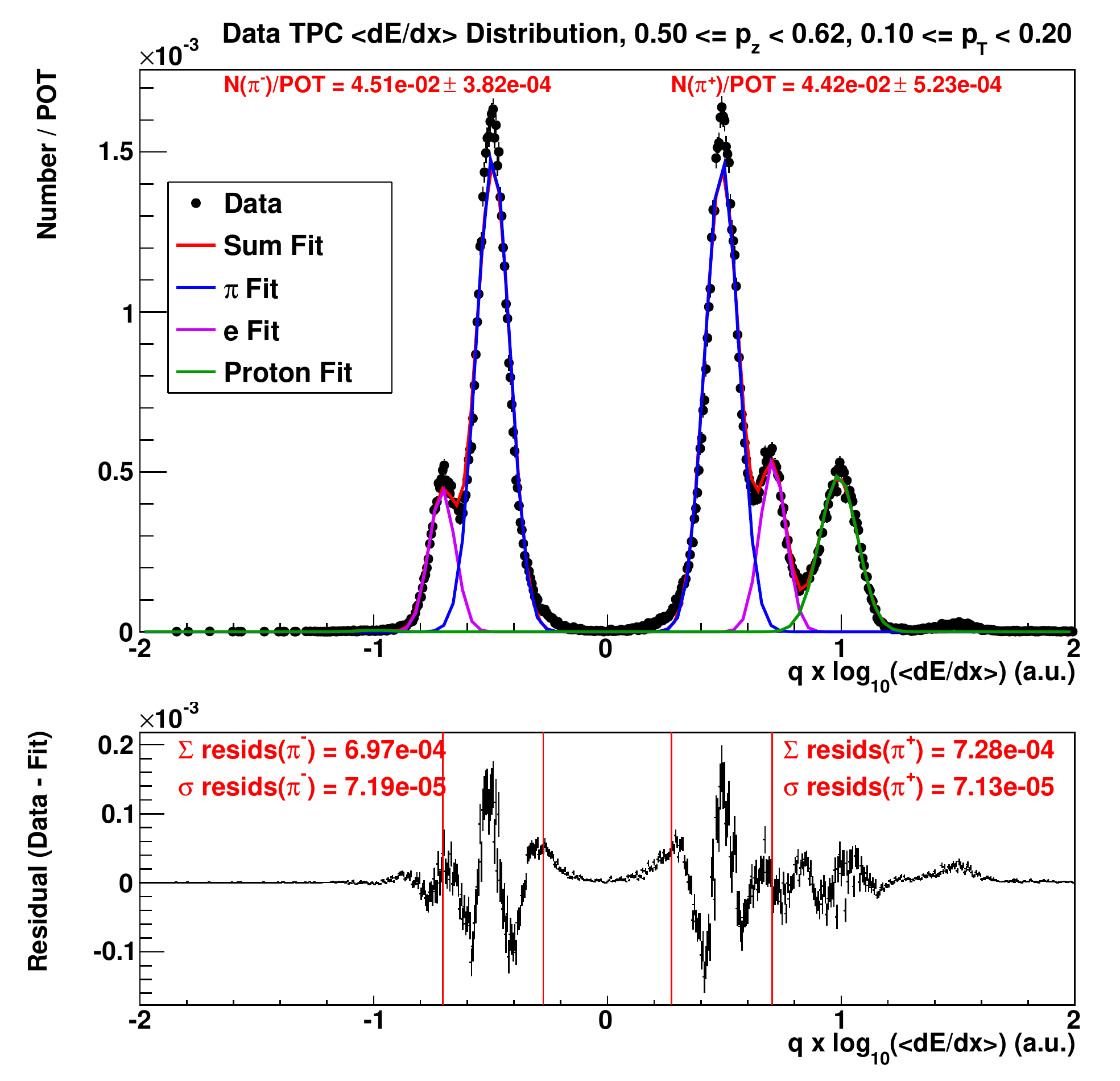}} 
  \subfigure[{$1.2 \le p_\mathrm{z} < 1.5, 0.20 \le p_\mathrm{T} < 0.30$ GeV/c.}]{\label{fig:fitdedx24} \includegraphics[width=0.45\textwidth]{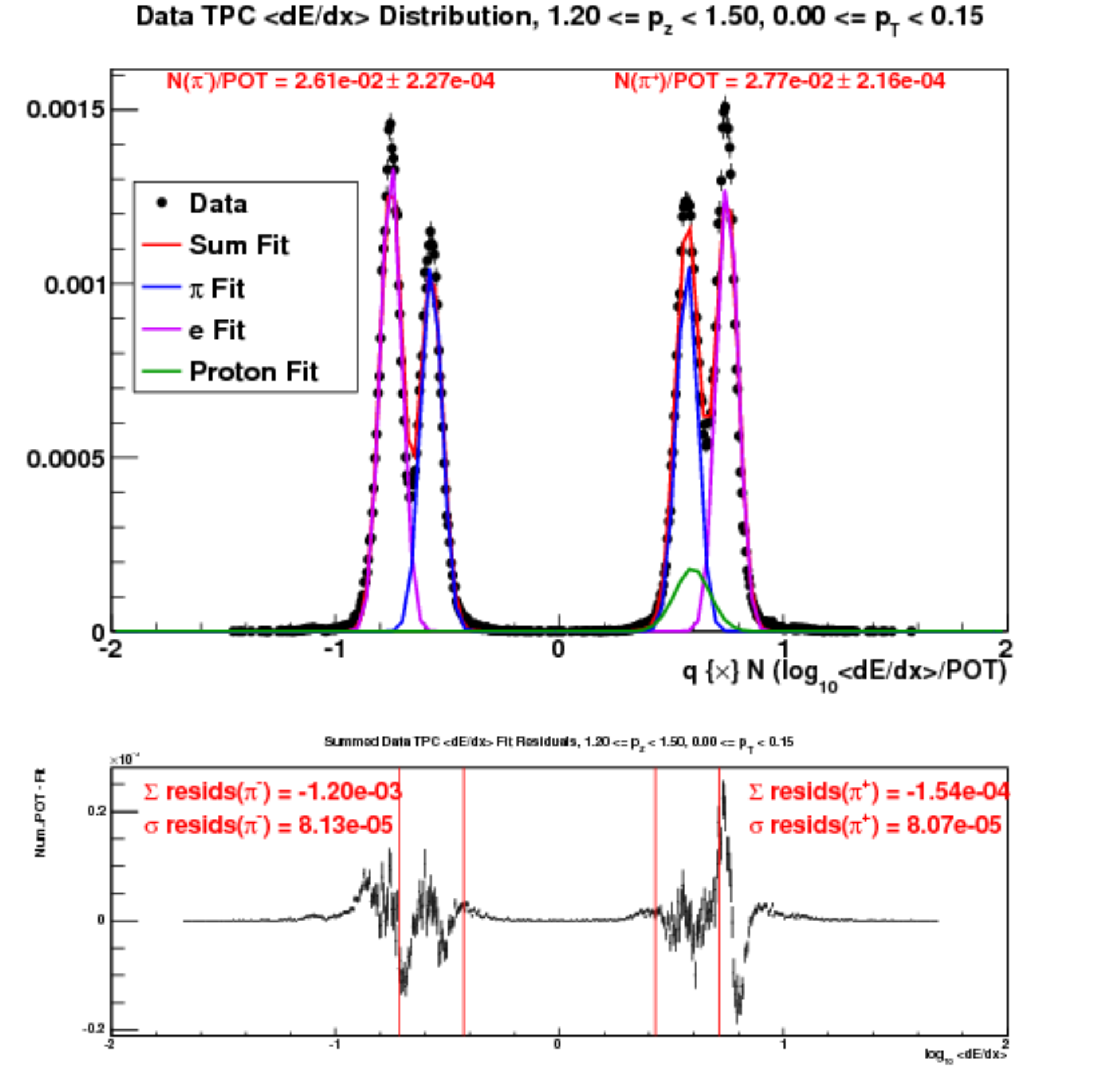}}%
     \caption{\dedx fit results to a sum of 6 Gaussians.  Residuals of the fit are shown below the fits.}
     \label{fig:dedxFits}
\end{figure*}

Figs.~\ref{fig:fitdedx7} and \ref{fig:fitdedx24} show two examples of fits to the \dedx distributions for two bins, the former at lower momentum where the proton peak is clearly visible, and the latter at higher momentum where the proton peak falls mostly under the pion peak.  In the latter case, the green curve is constrained from the ToF data as described above.  

The initial pion yield in each \pzpt bin is taken as the sum of the integrals of the fitted pion Gaussian peaks for the two independent data sets where ToF data are used in conjunction with TPC data, and where only TPC data are used.  The uncertainty on the pion yield in each case is taken from the uncertainty in the fit parameters for the amplitude and width.  However, it is clear that these fits are not perfect; the bottoms of Figs.~\ref{fig:fitdedx7} and \ref{fig:fitdedx24} show the residuals of the fit (data - fit).  We take into account the imperfection of the fit by adding the sum of the residuals in the range $[-3\sigma_\pi,3\sigma_\pi]$ (the red lines in the figures), where $\sigma_\pi$ is the fitted width of the pion Gaussian peak, and the RMS of the residuals in these regions is taken as the uncertainty on this correction to the pion yield.  These corrections are typically on the order of 10\%, and typical uncertainties of the corrections are on the order of 10\% thereby contributing less than 1\% to the total pion yield uncertainty.  All uncertainties are added in quadrature.

The TPC \dedx data show no clear kaon peak, and so we do not attempt to fit for the kaon contribution in Eq.~\ref{Eqn:dEdx}.  A kaon contribution is estimated using MC to predict the $\pi/K$ fraction in each bin; in most bins this contribution is well below 10\%.  The contribution is subtracted from the pion yield in each bin, and a 30\% uncertainty, which is an estimate of the level of uncertainty in the MC prediction of the particle yields off the NuMI target, is attributed to the correction.

\subsection{RICH Measurements}
\begin{figure*}[!htb] 
   \centering
   \subfigure[{} $6.0 \le p_\mathrm{z} < 8.0, 0.2 \le p_\mathrm{T} < 0.3$ GeV/c, negative particles.]{\label{fig:rich_count_minus_44}\includegraphics[width=0.45\textwidth]{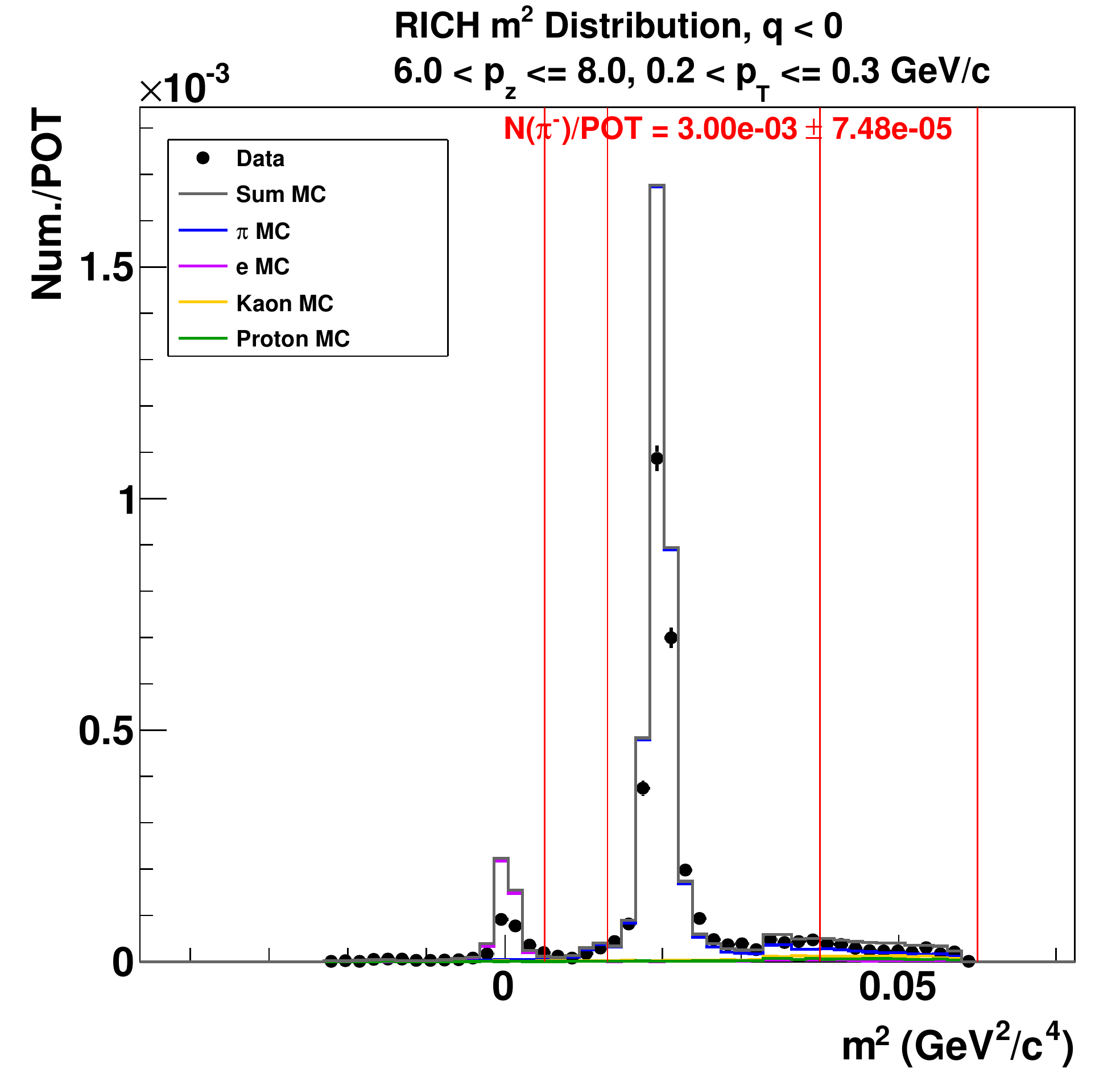}} 
   \subfigure[{} $12.0 \le p_\mathrm{z} < 14.0, 0.1 \le p_\mathrm{T} < 0.2$ GeV/c, positive particles.]{\label{fig:rich_count_minus_54}\includegraphics[width=0.45\textwidth]{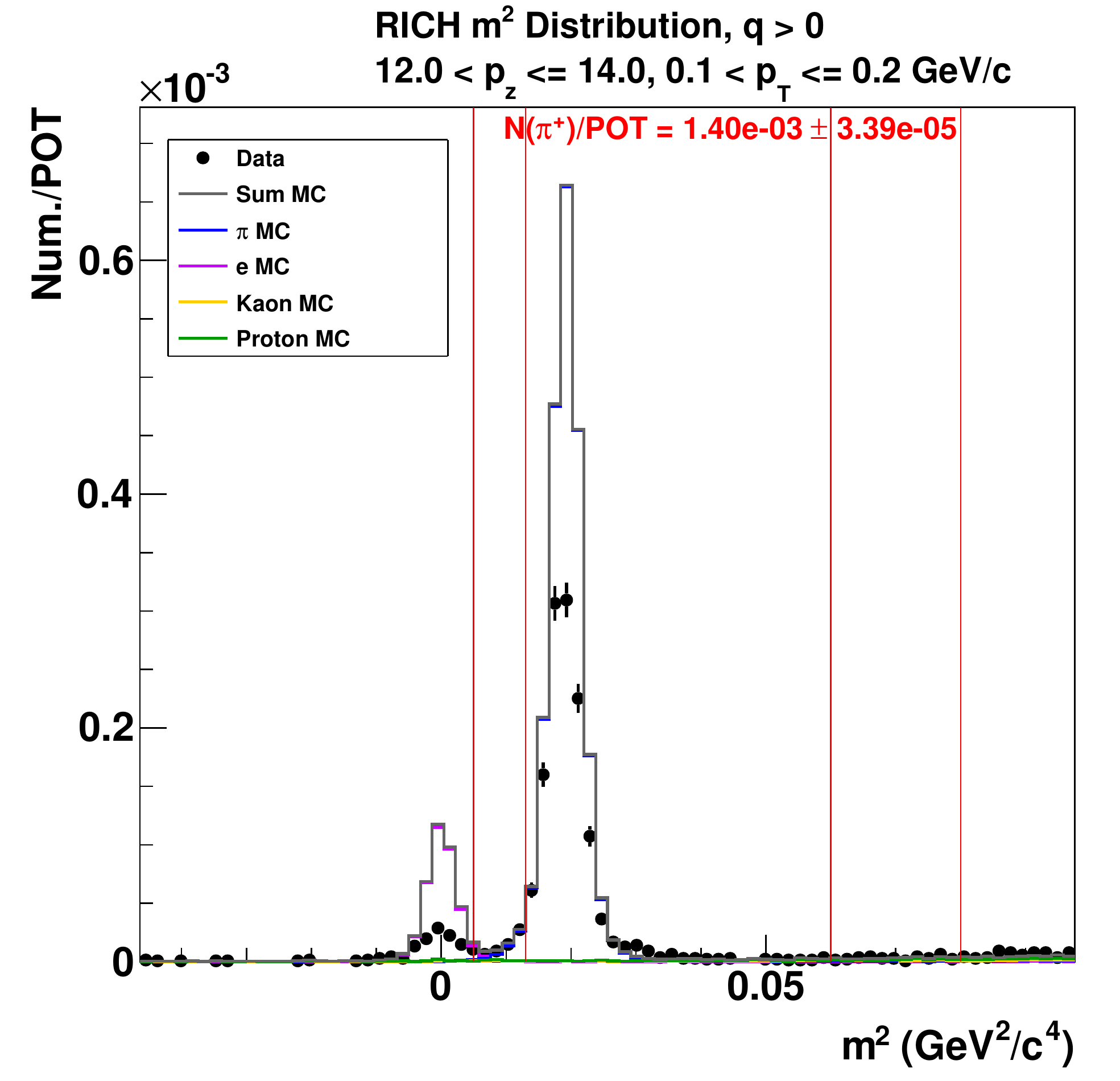}}%
   \caption{RICH $m^2$ distributions, data (black dots) vs. MC (solid lines).  The solid vertical red lines represent the bounds to define the signal and side-band windows for background estimation.}
   \label{fig:richm2A}
\end{figure*}
   
The RICH $m^2$ distributions are not well described by Gaussians; however, in general the $e, \pi, K$ and $p$ peaks in these distributions are quite well separated.  Therefore we take a simple cut-and-count approach, where we count the number of tracks that fall within a range in $m^2$ that contains pions.  In practice, however, there is contamination from non-pions, mostly electrons and positrons on the low-side tail of the pions, as well as some pion signal that sits under the electron peak, both of which must be taken into account.  We assume that the shapes of the $m^2$ distributions for each particle are properly modeled in the MC.  We then define three ranges, one main signal range and two side-band ranges.  The data in the sidebands are used to normalize the backgrounds predicted by the MC in the signal region.  Defining $N_i$ as the number of tracks within a range $i$, $\bar{N}_i$ as the number of tracks inside the other two ranges, $B_i$ as the MC background estimate (number of non-pions) in range $i$ and $\bar{S}_i$ [$\bar{B}_i$] as the MC signal [background] estimate inside the other two ranges, the pion yield is then
\begin{equation}
N(\pi) = \sum_i N_i(\pi),
\end{equation}
where
\begin{equation}
N_i(\pi) = N_i^\mathrm{Data} - b_i^\mathrm{MC}\bar{N}_i^\mathrm{Data},
\end{equation}
and
\begin{equation}
b_i = \frac{B_i}{\bar{S}_i + \bar{B}_i}.
\end{equation}

The uncertainty on the number of pions is

\begin{equation}
\sigma^2_{N(\pi)} = \sum_i \sigma^2_{N_i(\pi)}
\end{equation}

where 
\begin{equation}
\sigma^2_{N(\pi)_i} = N_i + \bar{N}_ib_i^2\left(1 + \bar{N}_i\left(\frac{\delta b_i}{b_i}\right)^2\right),
\end{equation}
$b_i$ represents the relative amounts of pion to non-pion background in each range in the MC, and we assume  a 30\% conservative systematic uncertainty in this ratio; the MC statistical uncertainty is negligible.  The positions and widths of the signal and side-band ranges are set by hand, based on a visual scan of the $m^2$ distribution in each bin.  The upper sideband where $m^2 > m^2_{\pi}$ contains almost no background.  In some cases, the MC predicts some small amount (few percent) of signal outside the ranges used to determine $N(\pi)$.  The predicted fraction of missing signal is added back to the measure $N(\pi)$ from data, and a 30\% uncertainty is attributed to this correction.  Fig.~\ref{fig:richm2A} shows examples of data and MC RICH $m^2$ distributions; the window boundaries are defined by the vertical solid red lines, and the pion yield and uncertainty are displayed near the top of each in red.  In most cases the RICH pion yield has uncertainties below 5\%.  The pion yield measurements depend very weakly on the exact positions of the sidebands.

\subsection{Statistical and Background Systematic Uncertainties}

\begin{figure*}[!htb] 
   \centering
   \subfigure[$\pi^+$]{\label{fig:StatSystPos}\includegraphics[width=0.45\textwidth]{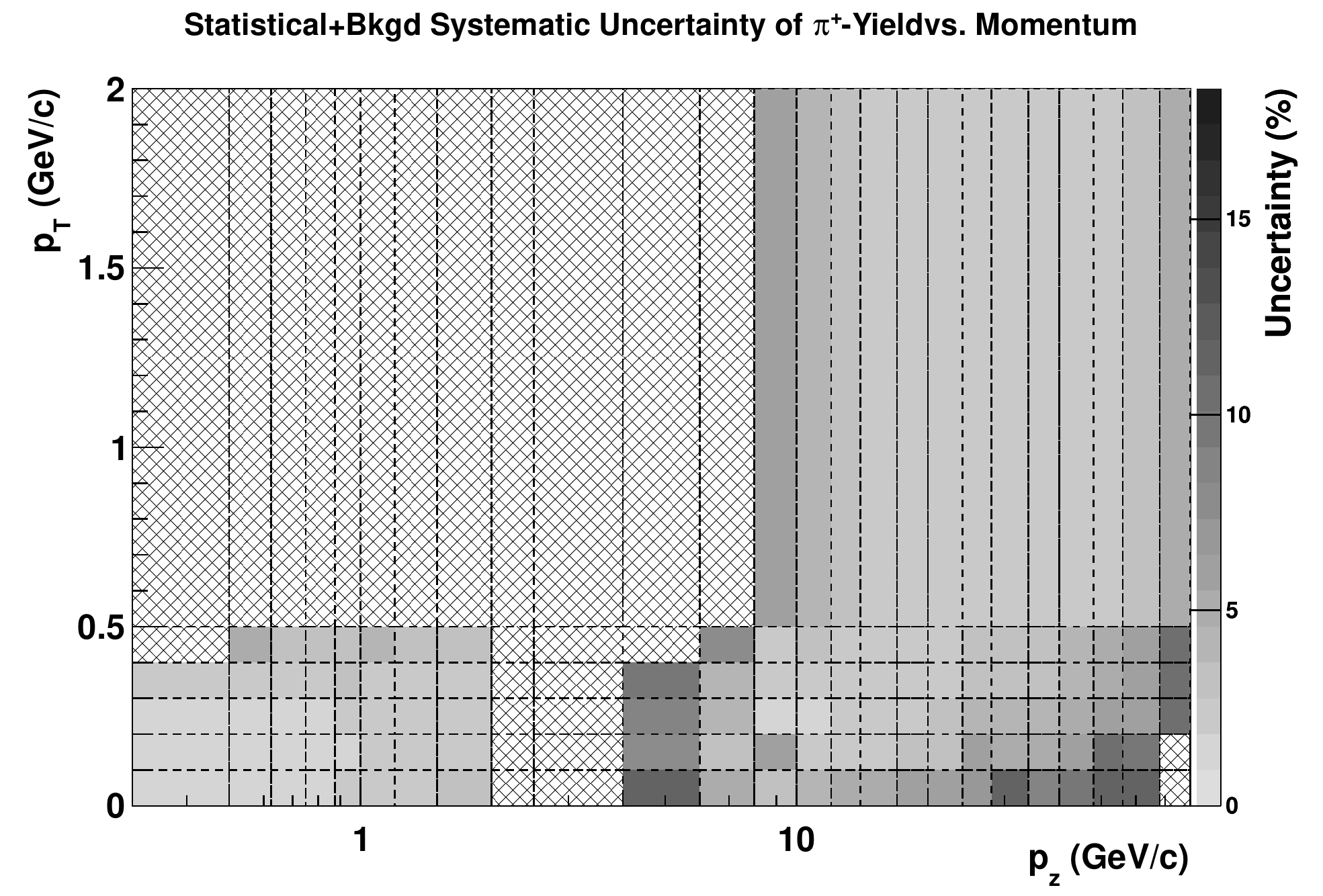}} 
   \subfigure[$\pi^-$]{\label{fig:StatSystNeg}\includegraphics[width=0.45\textwidth]{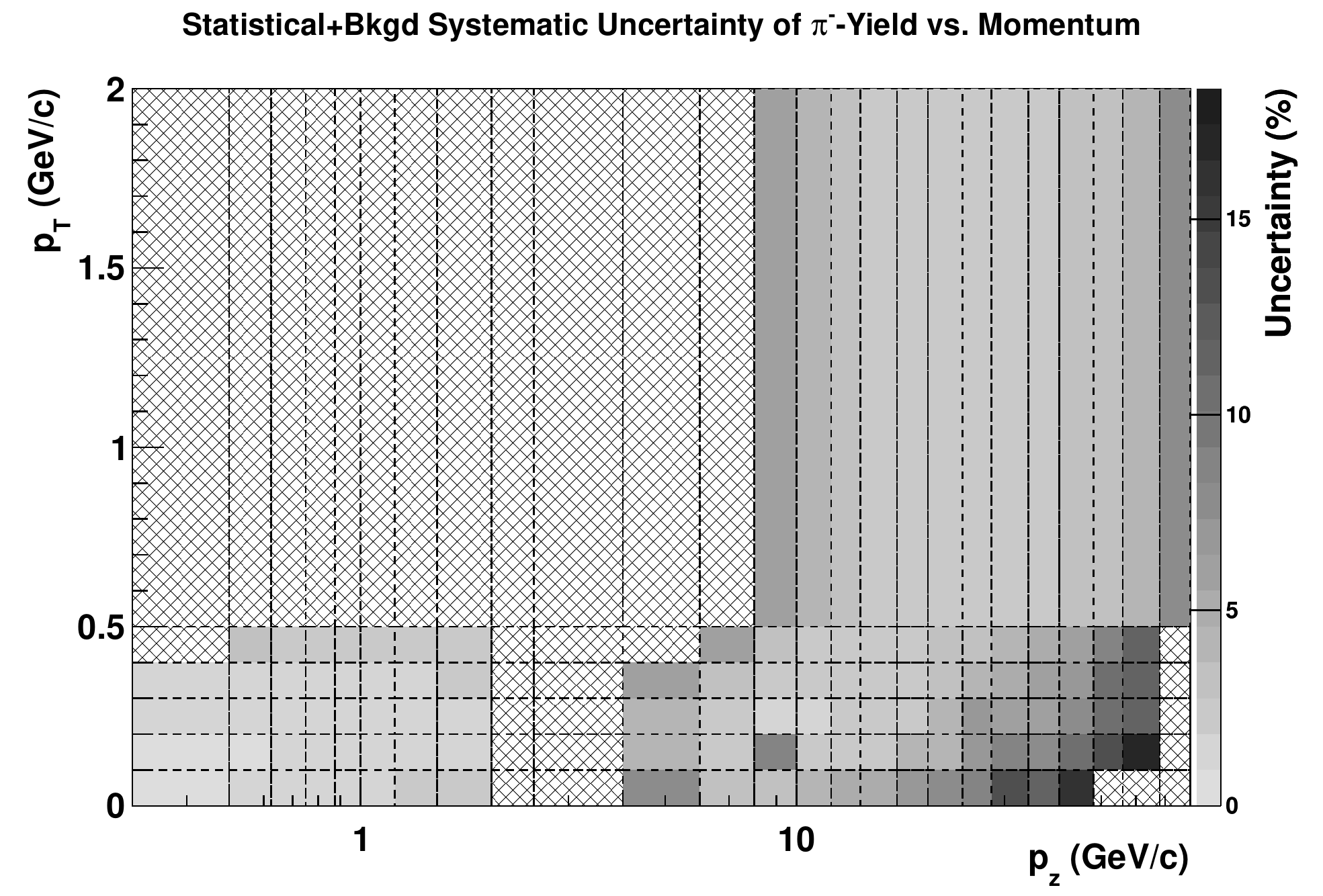}}%
\caption{Combined statistical and background systematic uncertainties of the charged pion yields as a function of \pzpt.  The shade of the bin represents the fractional uncertainty of each measurement.}
\label{fig:StatSyst}
\end{figure*}

The methods to determine the pion yields discussed above provide an 
uncertainty which combines statistical uncertainties and systematic uncertainties from backgrounds.  The relative 
uncertainties as a function of \pzpt are shown in Fig.~\ref{fig:StatSyst} for the $\pi^+$ yields; the uncertainties are similar for the $\pi^-$ yields.  In general, the combined uncertainty is a few percent for most bins of \pzpt where a measurement is made; the hashed bins are those excluded from the results because either no measurement was made or the uncertainty on the measurement is greater than 20\%.

\subsection{Systematic Uncertainties}
\begin{figure*}[!htb] 
   \centering
   \subfigure[$\pi^+$]{\label{fig:TotalSystPos}\includegraphics[width=0.45\textwidth]{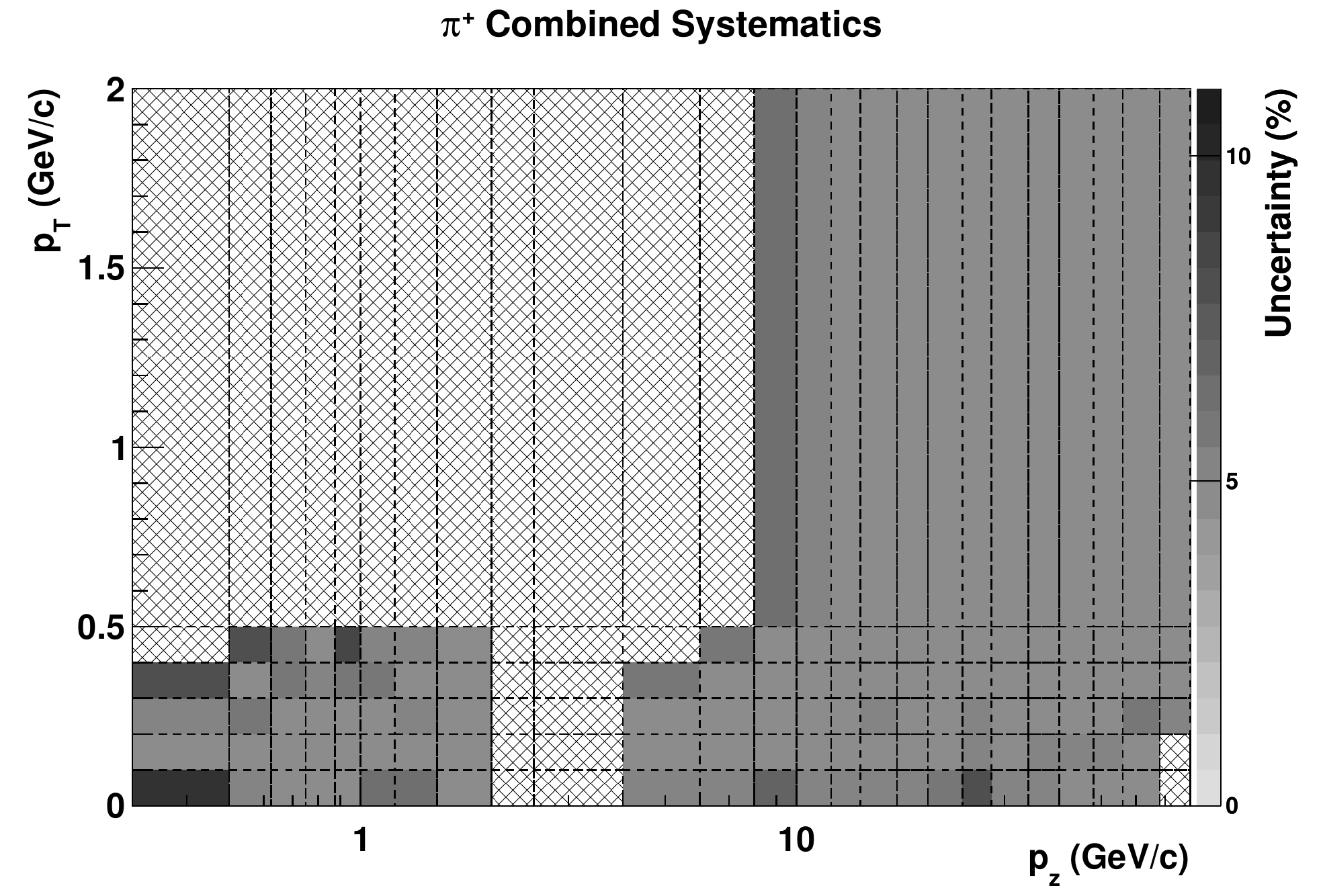}} 
   \subfigure[$\pi^-$]{\label{fig:TotalSystNeg}\includegraphics[width=0.45\textwidth]{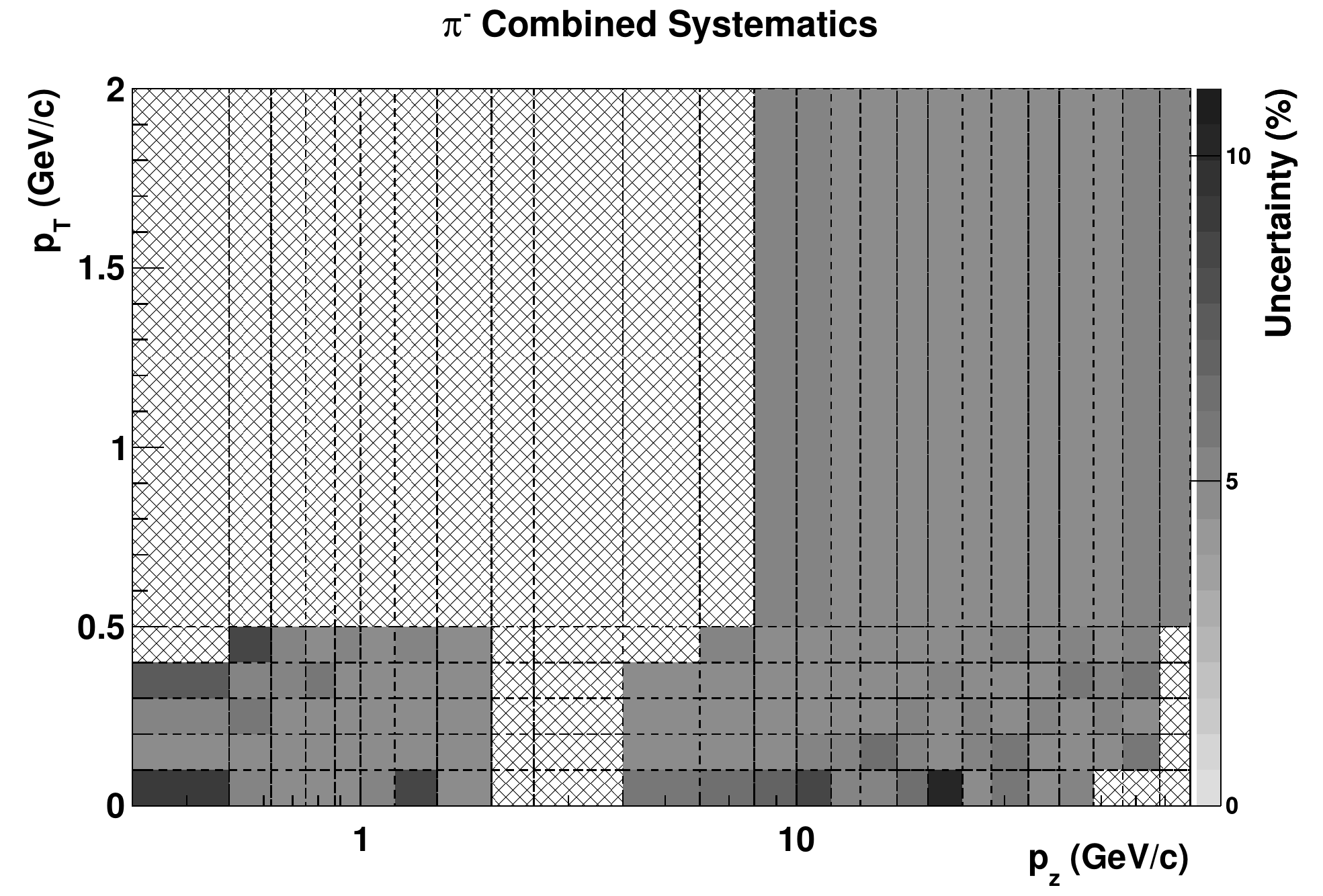}}%
\caption{Combined systematic uncertainties of the charged pion yields as a function of $p_\mathrm{z}$ in bins of $p_\mathrm{T}$.  The shade of the bin represents the fractional uncertainty of each measurement.}
\label{fig:TotalSystematics}
\end{figure*}

The momentum corrections discussed above are known to better than 10\%.  The systematic uncertainties due to a 10\% uncertainty on the momentum corrections were calculated for each \pzpt bin independently.  The effect is typically less than 1\%, with a few bins as large as 2.5\%.

Momentum resolution and reconstruction failures result in migrations of pions across bins of \pzpt.  Since the bins used in the analysis are much larger than the momentum resolution, the net effect of pion migrations across bins is small.  MC predictions imply this effect on the pion yield to be less than 4\% across all bins, which we take as the systematic uncertainty.  An additional systematic uncertainty of 1\% is added to account for mis-modeling of noise in the detectors which could result in discrepancies between the measured and predicted yields.  We note that these systematic effects do not cancel in the ratio of negative to positive pion yields.

Differences between the data and MC distributions of the GoF and DoCA variables are used to determine the track selection cut systematic uncertainties.  The GoF distributions agree to within 14\% in the relevant range of GoF variable values.  The GoF cuts were effectively varied by this $\pm 14\%$ in both data and MC and the impact on the measured pion yields determined in each \pzpt bin.  The relative difference between the measured pion yields using these modified GoF cuts and the nominal GoF cut is typically less than 2\%, with a  few lower momentum bins as large as 6\%.  A similar approach was taken for the DoCA cuts.  The data and MC DoCA distributions agree to within $\pm 10\%$, and we find that modifying these cuts in both data and MC results in a systematic shift of the pion yields of less than $1\%$ in nearly all bins, with a few lower momentum bins as large as 7\%.
 
The uncertainty of the acceptance and efficiency corrections that are applied to the measured yields in each bin arises from MC statistics ($< 1\%$), imperfections in the geometry model of the spectrometer ($< 1\%$), imperfections in the modeling of the time-dependent performance of the tracking and PID detectors, and incorrect modeling of the particle yields in the MC.  

The same time-dependency of thresholds used in the data collection, reconstruction and analysis was also applied to the reconstruction and analysis of the MC simulation data.  The time-dependent thresholds are known to within a few percent.  We therefore assume a 2\% uncertainty on all efficiencies in the pion yields due to imperfections in the modeling of the detectors in the MC.  This uncertainty effectively cancels out in the ratio of the pion yields.

Incorrect modeling of the particle yields in the MC results in improper modeling of the effect of overlapping secondary particles (pileup) in the tracking and PID detectors.  In high multiplicity events, the track reconstruction algorithm may either combine or confuse hits from different particles.  Pileup is the main cause of reconstruction and PID inefficiency; this effect can be as large as 30\% at low momenta, and is the main reason why results are not reported here for many \pzpt bins at lower momenta.  For bins where the pileup effect is below 20\%, MC events are re-weighted such that the multiplicity distribution (number of charged tracks coming off the surface of the target) in MC matches that of data, and the efficiencies are re-calculated.  We assume a conservative estimate of the uncertainty on the re-weighting factor to be 20\%, based on hand-scan studies that indicate the uncertainty on the measurement of event multiplicity is much less than 20\%.  We therefore take 20\% of the relative difference between the efficiencies determined from the nominal and the re-weighted MC as the systematic uncertainty due to this effect.  Typical relative uncertainties are a few percent, although some bins have uncertainties as high as 10\%.

The systematic uncertainties in the pion yield measurement described above are added in quadrature and displayed as a function of \pzpt in Fig.~\ref{fig:TotalSystematics} for both positive and negative pions.  Nearly all bins have systematic uncertainties between 4 and 5\%.

\subsection{Results}

\begin{figure*}[!htb] 
   \centering
   \subfigure[$\pi^+$ yields.]{\label{fig:FinalPiPlusYields}\includegraphics[width=0.4\textwidth]{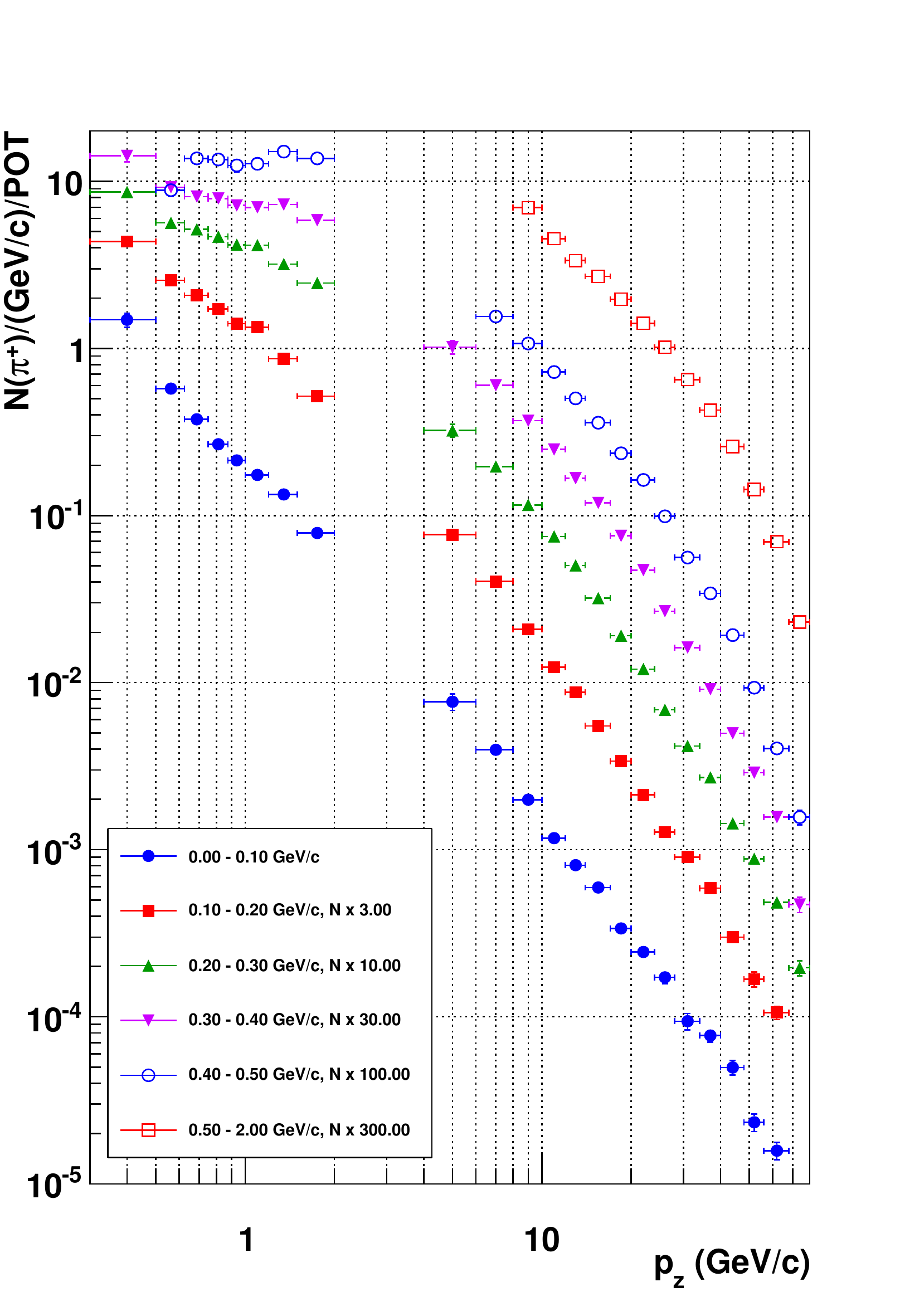}} 
   \subfigure[$\pi^-$ yields.]{\label{fig:FinalPiMinusYields}\includegraphics[width=0.4\textwidth]{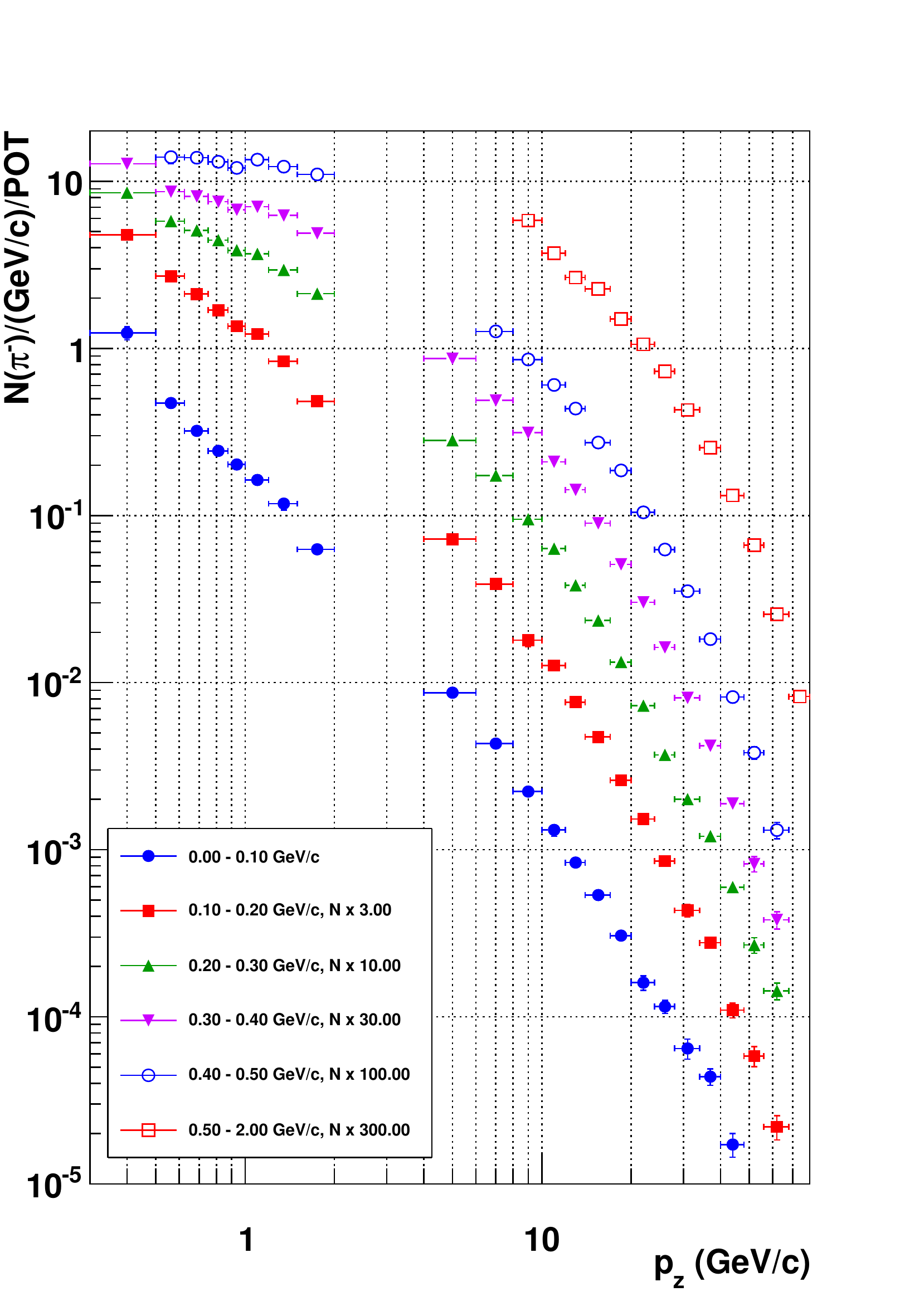}}
   \subfigure[$\pi^-/\pi^+$ ratios.]{\label{fig:FinalPiRatios}\includegraphics[width=0.4\textwidth]{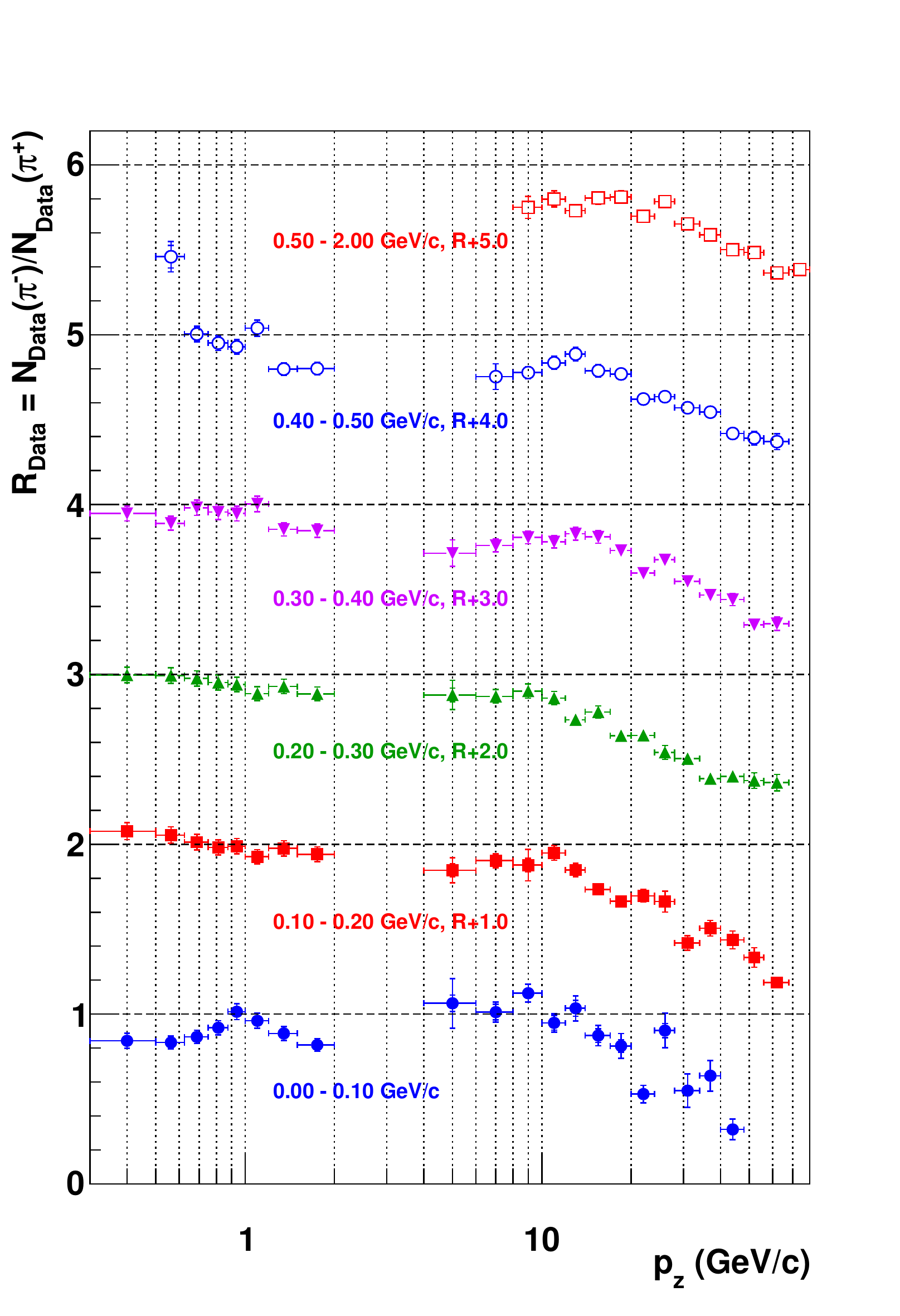}}%
\caption{Pion yields (top) and ratios (bottom) as a function of $p_\mathrm{z}$ in bins of $p_\mathrm{T}$.  Different colors and markers represent bins of $p_\mathrm{T}$, and the yields are scaled and the ratios offset such that the points in different $p_\mathrm{T}$ bins do not overlap.  All efficiency corrections have been applied, and both statistical and systematic error bars are plotted.}
\label{fig:FinalPionYields}
\end{figure*}

The measured $N(\pi^+)$/POT and $N(\pi^-)$/POT per \pzpt bin, along with the combined statistical and systematic errors, are shown in Figs.~\ref{fig:FinalPiPlusYields} and \ref{fig:FinalPiMinusYields} and Table \ref{table:FinalPionYields}.  The data points in the plots in Figs.~\ref{fig:FinalPiPlusYields} and \ref{fig:FinalPiMinusYields} have been normalized by the width of the momentum bins.  The uncertainties in the table are in units of percent.  We see that in most of the bins, the measurements are systematics limited and nearly all measurements have uncertainties estimated below 10\%.  Fig.~\ref{fig:FinalPiRatios} shows the ratio, $R$, of
 $\pi^-/\pi^+$ yields as a function of $p_\mathrm{z}$ in slices of $p_\mathrm{T}$.  Table \ref{table:FinalPionYields} also lists the $R$-values measured in each bin along with the statistical and systematic uncertainties.  Correlated systematics between the positive and negative pion yields have been canceled out in the ratios.

\clearpage

\section{Summary}
A measurement of $\pi^+$ and $\pi^-$ yields from 120 GeV/c protons incident on a NuMI low-energy target has been performed across 124 and 119 bins of \pzpt bins respectively using data collected in the MIPP fixed-target experiment at Fermilab.  Typical uncertainties on the measurements in each bin are a few percent.  These data may be directly used to improve the calculation and the uncertainties on the calculation of the neutrino flux in the NuMI beam line.

\section{Acknowledgements}
This work was supported by the US Department of Energy.  We are grateful to the staff of Fermilab, Lawrence Livermore National Laboratory under Contract DE-AC52-07NA27344 and Argonne National Laboratory for their contributions to this effort.

\newpage

\begin{longtable*}{|c|c|c|c|c|c|c|c|c|c|c|}
\caption{Pion Production Yields - NuMI target}
\label{table:FinalPionYields}\\

\hline
& & &\textbf{$\delta N(\pi^+)$} &\textbf{$\delta N(\pi^+)$} & &\textbf{$\delta N(\pi^-)$} &\textbf{$\delta N(\pi^-)$} &\textbf{$R = $} & \textbf{$\delta R$}& \textbf{$\delta R$} \\
\textbf{$p_\mathrm{z}$} & \textbf{$p_\mathrm{T}$} & \textbf{$N(\pi^+)/$} & \textbf{stat+} & \textbf{syst} & \textbf{$N(\pi^-)/$} & \textbf{stat+} & \textbf{syst} & \textbf{$N(\pi^-)/$} & \textbf{stat+} & \textbf{syst} \\
\textbf{(GeV/c)} &\textbf{(GeV/c)} & \textbf{POT}& \textbf{bkgd}& \textbf{(\%)}& \textbf{POT}& \textbf{bkgd}& \textbf{(\%)} & \textbf{$N(\pi^+)$}& \textbf{bkgd}& \textbf{(\%)}\\
& & & \textbf{(\%)} & & & \textbf{(\%)} & & & \textbf{(\%)} & \\

\endfirsthead

\multicolumn{11}{c}{{\tablename} \thetable{} -- $\pi^+$ Yields, continued from previous page} \\
\endhead

\multicolumn{11}{|r|}{{Continued on next page\dots}} \\ 
\endfoot

\hline
\hline
\endlastfoot

\hline
[0.30,0.50) & [0.00,0.10) & 3.32e-01 & 0.97 & 9.77 & 2.76e-01 & 0.86 & 9.09 & 0.84 & 1.29 & 5.19 \\ 
 \hline
[0.30,0.50) & [0.10,0.20) & 3.25e-01 & 1.05 & 4.86 & 3.56e-01 & 0.75 & 4.93 & 1.08 & 1.29 & 4.64 \\ 
 \hline
[0.30,0.50) & [0.20,0.30) & 1.93e-01 & 1.46 & 5.30 & 1.91e-01 & 1.09 & 5.24 & 1.00 & 1.82 & 4.59 \\ 
 \hline
[0.30,0.50) & [0.30,0.40) & 1.06e-01 & 1.97 & 8.15 & 9.49e-02 & 1.68 & 7.39 & 0.95 & 2.59 & 4.61 \\ 
 \hline
[0.50,0.62) & [0.00,0.10) & 9.20e-02 & 1.22 & 5.10 & 7.54e-02 & 1.16 & 4.98 & 0.83 & 1.69 & 4.59 \\ 
 \hline
[0.50,0.62) & [0.10,0.20) & 1.37e-01 & 1.18 & 4.72 & 1.44e-01 & 0.85 & 4.70 & 1.05 & 1.46 & 4.59 \\ 
 \hline
[0.50,0.62) & [0.20,0.30) & 9.03e-02 & 1.57 & 5.64 & 9.24e-02 & 1.14 & 5.92 & 0.99 & 1.94 & 4.59 \\ 
 \hline
[0.50,0.62) & [0.30,0.40) & 4.92e-02 & 2.12 & 4.77 & 4.63e-02 & 1.76 & 5.24 & 0.89 & 2.76 & 4.58 \\ 
 \hline
[0.50,0.62) & [0.40,0.50) & 1.42e-02 & 5.39 & 8.18 & 2.23e-02 & 2.81 & 8.33 & 1.46 & 6.08 & 4.58 \\ 
 \hline
[0.62,0.75) & [0.00,0.10) & 6.02e-02 & 1.41 & 4.94 & 5.14e-02 & 1.31 & 4.85 & 0.87 & 1.93 & 4.59 \\ 
 \hline
[0.62,0.75) & [0.10,0.20) & 1.11e-01 & 1.33 & 4.75 & 1.13e-01 & 0.94 & 4.74 & 1.01 & 1.63 & 4.58 \\ 
 \hline
[0.62,0.75) & [0.20,0.30) & 8.26e-02 & 1.59 & 4.65 & 8.14e-02 & 1.12 & 4.63 & 0.98 & 1.94 & 4.58 \\ 
 \hline
[0.62,0.75) & [0.30,0.40) & 4.32e-02 & 2.09 & 5.62 & 4.33e-02 & 1.52 & 5.43 & 0.98 & 2.58 & 4.58 \\ 
 \hline
[0.62,0.75) & [0.40,0.50) & 2.19e-02 & 3.12 & 5.78 & 2.22e-02 & 2.47 & 5.47 & 1.00 & 3.98 & 4.58 \\ 
 \hline
[0.75,0.88) & [0.00,0.10) & 4.27e-02 & 1.74 & 4.74 & 3.90e-02 & 1.55 & 4.68 & 0.92 & 2.33 & 4.58 \\ 
 \hline
[0.75,0.88) & [0.10,0.20) & 9.19e-02 & 1.48 & 4.68 & 9.04e-02 & 1.05 & 4.65 & 0.98 & 1.82 & 4.58 \\ 
 \hline
[0.75,0.88) & [0.20,0.30) & 7.46e-02 & 1.67 & 4.62 & 7.11e-02 & 1.18 & 4.61 & 0.95 & 2.05 & 4.58 \\ 
 \hline
[0.75,0.88) & [0.30,0.40) & 4.21e-02 & 2.19 & 5.50 & 4.03e-02 & 1.54 & 5.68 & 0.96 & 2.68 & 4.58 \\ 
 \hline
[0.75,0.88) & [0.40,0.50) & 2.16e-02 & 2.84 & 4.71 & 2.09e-02 & 2.26 & 4.77 & 0.95 & 3.63 & 4.58 \\ 
 \hline
[0.88,1.00) & [0.00,0.10) & 3.42e-02 & 1.87 & 4.74 & 3.23e-02 & 1.62 & 4.69 & 1.01 & 2.48 & 4.58 \\ 
 \hline
[0.88,1.00) & [0.10,0.20) & 7.50e-02 & 1.69 & 4.68 & 7.25e-02 & 1.22 & 4.62 & 0.99 & 2.09 & 4.58 \\ 
 \hline
[0.88,1.00) & [0.20,0.30) & 6.68e-02 & 1.85 & 4.64 & 6.19e-02 & 1.33 & 4.60 & 0.94 & 2.28 & 4.58 \\ 
 \hline
[0.88,1.00) & [0.30,0.40) & 3.82e-02 & 2.87 & 5.57 & 3.61e-02 & 1.58 & 4.68 & 0.95 & 3.27 & 4.58 \\ 
 \hline
[0.88,1.00) & [0.40,0.50) & 1.99e-02 & 2.90 & 8.53 & 1.93e-02 & 2.15 & 5.04 & 0.93 & 3.61 & 4.59 \\ 
 \hline
[1.00,1.20) & [0.00,0.10) & 3.91e-02 & 2.04 & 6.21 & 3.65e-02 & 1.71 & 5.33 & 0.96 & 2.66 & 4.59 \\ 
 \hline
[1.00,1.20) & [0.10,0.20) & 9.98e-02 & 1.83 & 4.65 & 9.09e-02 & 1.31 & 4.64 & 0.93 & 2.25 & 4.59 \\ 
 \hline
[1.00,1.20) & [0.20,0.30) & 9.27e-02 & 2.06 & 4.62 & 8.22e-02 & 1.41 & 4.60 & 0.89 & 2.50 & 4.59 \\ 
 \hline
[1.00,1.20) & [0.30,0.40) & 5.20e-02 & 2.46 & 5.65 & 5.25e-02 & 1.67 & 4.74 & 1.00 & 2.97 & 4.59 \\ 
 \hline
[1.00,1.20) & [0.40,0.50) & 2.85e-02 & 4.18 & 5.42 & 3.01e-02 & 2.05 & 4.63 & 1.04 & 4.65 & 4.58 \\ 
 \hline
[1.20,1.50) & [0.00,0.10) & 4.23e-02 & 2.35 & 6.27 & 3.72e-02 & 1.91 & 8.47 & 0.89 & 3.02 & 4.59 \\ 
 \hline
[1.20,1.50) & [0.10,0.20) & 9.12e-02 & 2.07 & 4.61 & 8.83e-02 & 1.50 & 4.62 & 0.98 & 2.56 & 4.58 \\ 
 \hline
[1.20,1.50) & [0.20,0.30) & 1.01e-01 & 2.20 & 5.00 & 9.32e-02 & 1.60 & 4.90 & 0.93 & 2.72 & 4.59 \\ 
 \hline
[1.20,1.50) & [0.30,0.40) & 7.65e-02 & 2.51 & 4.64 & 6.58e-02 & 1.80 & 4.60 & 0.85 & 3.08 & 4.59 \\ 
 \hline
[1.20,1.50) & [0.40,0.50) & 4.76e-02 & 3.11 & 5.25 & 3.87e-02 & 2.19 & 5.40 & 0.80 & 3.80 & 4.59 \\ 
 \hline
[1.50,2.00) & [0.00,0.10) & 4.01e-02 & 2.90 & 4.62 & 3.20e-02 & 2.58 & 4.59 & 0.82 & 3.88 & 4.58 \\ 
 \hline
[1.50,2.00) & [0.10,0.20) & 8.80e-02 & 2.44 & 4.88 & 8.20e-02 & 1.79 & 4.93 & 0.94 & 3.03 & 4.59 \\ 
 \hline
[1.50,2.00) & [0.20,0.30) & 1.26e-01 & 2.41 & 4.65 & 1.08e-01 & 1.73 & 4.65 & 0.89 & 2.97 & 4.59 \\ 
 \hline
[1.50,2.00) & [0.30,0.40) & 9.94e-02 & 2.70 & 4.71 & 8.30e-02 & 1.95 & 4.66 & 0.85 & 3.33 & 4.59 \\ 
 \hline
[1.50,2.00) & [0.40,0.50) & 6.99e-02 & 3.14 & 4.63 & 5.61e-02 & 2.31 & 4.63 & 0.80 & 3.90 & 4.59 \\ 
 \hline
[4.00,6.00) & [0.00,0.10) & 1.54e-02 & 11.19 & 5.03 & 1.74e-02 & 7.94 & 5.83 & 1.06 & 13.73 & 4.58 \\ 
 \hline
[4.00,6.00) & [0.10,0.20) & 5.12e-02 & 7.63 & 4.67 & 4.82e-02 & 4.23 & 4.91 & 0.85 & 8.72 & 4.58 \\ 
 \hline
[4.00,6.00) & [0.20,0.30) & 6.47e-02 & 8.72 & 4.79 & 5.64e-02 & 4.48 & 4.86 & 0.88 & 9.81 & 4.59 \\ 
 \hline
[4.00,6.00) & [0.30,0.40) & 6.79e-02 & 9.31 & 5.61 & 5.79e-02 & 5.64 & 4.73 & 0.71 & 10.89 & 4.60 \\ 
 \hline
[6.00,8.00) & [0.00,0.10) & 7.94e-03 & 4.78 & 5.40 & 8.64e-03 & 3.46 & 6.40 & 1.01 & 5.90 & 4.58 \\ 
 \hline
[6.00,8.00) & [0.10,0.20) & 2.69e-02 & 3.55 & 4.64 & 2.60e-02 & 2.57 & 4.95 & 0.90 & 4.38 & 4.58 \\ 
 \hline
[6.00,8.00) & [0.20,0.30) & 3.93e-02 & 3.82 & 4.69 & 3.47e-02 & 2.49 & 4.62 & 0.87 & 4.56 & 4.58 \\ 
 \hline
[6.00,8.00) & [0.30,0.40) & 4.01e-02 & 3.88 & 4.79 & 3.26e-02 & 3.14 & 4.71 & 0.76 & 4.99 & 4.59 \\ 
 \hline
[6.00,8.00) & [0.40,0.50) & 3.11e-02 & 7.89 & 5.92 & 2.53e-02 & 6.11 & 5.24 & 0.75 & 9.98 & 4.59 \\ 
 \hline
[8.00,10.00) & [0.00,0.10) & 3.99e-03 & 3.55 & 6.90 & 4.46e-03 & 3.22 & 7.02 & 1.12 & 4.79 & 4.58 \\ 
 \hline
[8.00,10.00) & [0.10,0.20) & 1.39e-02 & 6.28 & 4.89 & 1.20e-02 & 8.47 & 4.84 & 0.88 & 10.54 & 4.58 \\ 
 \hline
[8.00,10.00) & [0.20,0.30) & 2.32e-02 & 1.70 & 4.69 & 1.90e-02 & 1.60 & 4.73 & 0.90 & 2.34 & 4.58 \\ 
 \hline
[8.00,10.00) & [0.30,0.40) & 2.46e-02 & 1.86 & 4.66 & 2.09e-02 & 1.91 & 4.68 & 0.81 & 2.67 & 4.58 \\ 
 \hline
[8.00,10.00) & [0.40,0.50) & 2.14e-02 & 2.64 & 4.66 & 1.72e-02 & 2.87 & 4.68 & 0.78 & 3.90 & 4.58 \\ 
 \hline
[8.00,10.00) & [0.50,2.00) & 5.80e-02 & 5.87 & 6.20 & 4.86e-02 & 6.33 & 5.19 & 0.75 & 8.63 & 4.60 \\ 
 \hline
[10.00,12.00) & [0.00,0.10) & 2.35e-03 & 4.01 & 5.30 & 2.63e-03 & 4.12 & 8.46 & 0.95 & 5.75 & 4.58 \\ 
 \hline
[10.00,12.00) & [0.10,0.20) & 8.25e-03 & 1.95 & 4.67 & 8.44e-03 & 2.01 & 4.97 & 0.95 & 2.80 & 4.58 \\ 
 \hline
[10.00,12.00) & [0.20,0.30) & 1.50e-02 & 1.72 & 4.63 & 1.27e-02 & 1.78 & 4.62 & 0.86 & 2.47 & 4.58 \\ 
 \hline
[10.00,12.00) & [0.30,0.40) & 1.66e-02 & 2.10 & 4.75 & 1.40e-02 & 1.91 & 4.69 & 0.78 & 2.84 & 4.58 \\ 
 \hline
[10.00,12.00) & [0.40,0.50) & 1.45e-02 & 2.79 & 4.70 & 1.21e-02 & 2.62 & 4.80 & 0.83 & 3.83 & 4.58 \\ 
 \hline
[10.00,12.00) & [0.50,2.00) & 3.78e-02 & 4.18 & 5.46 & 3.10e-02 & 4.26 & 5.17 & 0.80 & 5.97 & 4.59 \\ 
 \hline
[12.00,14.00) & [0.00,0.10) & 1.62e-03 & 4.98 & 5.34 & 1.68e-03 & 5.10 & 5.28 & 1.03 & 7.12 & 4.58 \\ 
 \hline
[12.00,14.00) & [0.10,0.20) & 5.84e-03 & 2.43 & 4.89 & 5.10e-03 & 2.61 & 5.08 & 0.85 & 3.56 & 4.58 \\ 
 \hline
[12.00,14.00) & [0.20,0.30) & 1.01e-02 & 2.00 & 4.62 & 7.65e-03 & 2.11 & 4.72 & 0.73 & 2.91 & 4.58 \\ 
 \hline
[12.00,14.00) & [0.30,0.40) & 1.12e-02 & 2.14 & 4.65 & 9.51e-03 & 2.12 & 4.61 & 0.83 & 3.02 & 4.58 \\ 
 \hline
[12.00,14.00) & [0.40,0.50) & 1.01e-02 & 2.40 & 4.68 & 8.74e-03 & 2.38 & 4.74 & 0.89 & 3.38 & 4.58 \\ 
 \hline
[12.00,14.00) & [0.50,2.00) & 2.79e-02 & 2.96 & 4.72 & 2.21e-02 & 2.96 & 4.70 & 0.73 & 4.19 & 4.58 \\ 
 \hline
[14.00,17.00) & [0.00,0.10) & 1.78e-03 & 4.62 & 5.19 & 1.61e-03 & 5.05 & 5.45 & 0.87 & 6.85 & 4.58 \\ 
 \hline
[14.00,17.00) & [0.10,0.20) & 5.51e-03 & 2.50 & 4.78 & 4.73e-03 & 2.68 & 6.56 & 0.73 & 3.66 & 4.58 \\ 
 \hline
[14.00,17.00) & [0.20,0.30) & 9.61e-03 & 1.88 & 5.01 & 7.06e-03 & 2.09 & 4.68 & 0.78 & 2.81 & 4.58 \\ 
 \hline
[14.00,17.00) & [0.30,0.40) & 1.19e-02 & 1.87 & 4.72 & 8.99e-03 & 2.01 & 4.69 & 0.81 & 2.74 & 4.58 \\ 
 \hline
[14.00,17.00) & [0.40,0.50) & 1.08e-02 & 2.12 & 4.64 & 8.20e-03 & 2.27 & 4.72 & 0.79 & 3.10 & 4.58 \\ 
 \hline
[14.00,17.00) & [0.50,2.00) & 3.02e-02 & 2.15 & 4.66 & 2.54e-02 & 2.01 & 4.61 & 0.81 & 2.95 & 4.58 \\ 
 \hline
[17.00,20.00) & [0.00,0.10) & 1.01e-03 & 5.78 & 5.12 & 9.19e-04 & 6.85 & 6.20 & 0.81 & 8.96 & 4.58 \\ 
 \hline
[17.00,20.00) & [0.10,0.20) & 3.39e-03 & 2.99 & 4.71 & 2.61e-03 & 3.66 & 5.45 & 0.66 & 4.73 & 4.58 \\ 
 \hline
[17.00,20.00) & [0.20,0.30) & 5.73e-03 & 2.17 & 4.65 & 3.98e-03 & 2.68 & 5.10 & 0.64 & 3.45 & 4.58 \\ 
 \hline
[17.00,20.00) & [0.30,0.40) & 7.56e-03 & 2.45 & 4.74 & 5.11e-03 & 2.73 & 4.72 & 0.73 & 3.67 & 4.58 \\ 
 \hline
[17.00,20.00) & [0.40,0.50) & 7.07e-03 & 2.55 & 4.70 & 5.59e-03 & 2.69 & 4.71 & 0.77 & 3.71 & 4.58 \\ 
 \hline
[17.00,20.00) & [0.50,2.00) & 2.20e-02 & 1.88 & 4.62 & 1.68e-02 & 2.01 & 4.69 & 0.81 & 2.75 & 4.58 \\ 
 \hline
[20.00,24.00) & [0.00,0.10) & 9.77e-04 & 5.79 & 5.85 & 6.40e-04 & 7.74 & 10.03 & 0.53 & 9.66 & 4.58 \\ 
 \hline
[20.00,24.00) & [0.10,0.20) & 2.84e-03 & 3.22 & 4.73 & 2.04e-03 & 4.12 & 4.66 & 0.70 & 5.23 & 4.58 \\ 
 \hline
[20.00,24.00) & [0.20,0.30) & 4.82e-03 & 2.98 & 4.89 & 2.91e-03 & 3.87 & 4.67 & 0.64 & 4.89 & 4.58 \\ 
 \hline
[20.00,24.00) & [0.30,0.40) & 6.28e-03 & 2.52 & 4.67 & 4.04e-03 & 3.10 & 5.05 & 0.60 & 4.00 & 4.58 \\ 
 \hline
[20.00,24.00) & [0.40,0.50) & 6.53e-03 & 2.72 & 4.66 & 4.18e-03 & 2.95 & 4.64 & 0.62 & 4.01 & 4.58 \\ 
 \hline
[20.00,24.00) & [0.50,2.00) & 2.01e-02 & 2.02 & 4.75 & 1.51e-02 & 1.89 & 4.65 & 0.70 & 2.77 & 4.58 \\ 
 \hline
[24.00,28.00) & [0.00,0.10) & 6.89e-04 & 6.65 & 8.24 & 4.61e-04 & 9.09 & 4.68 & 0.90 & 11.27 & 4.58 \\ 
 \hline
[24.00,28.00) & [0.10,0.20) & 1.70e-03 & 5.74 & 5.06 & 1.14e-03 & 7.25 & 5.09 & 0.66 & 9.25 & 4.58 \\ 
 \hline
[24.00,28.00) & [0.20,0.30) & 2.75e-03 & 3.92 & 5.28 & 1.48e-03 & 6.47 & 5.29 & 0.54 & 7.56 & 4.58 \\ 
 \hline
[24.00,28.00) & [0.30,0.40) & 3.57e-03 & 3.49 & 4.73 & 2.17e-03 & 3.92 & 4.78 & 0.68 & 5.25 & 4.58 \\ 
 \hline
[24.00,28.00) & [0.40,0.50) & 3.96e-03 & 3.35 & 4.82 & 2.50e-03 & 3.49 & 4.67 & 0.64 & 4.83 & 4.58 \\ 
 \hline
[24.00,28.00) & [0.50,2.00) & 1.45e-02 & 2.05 & 4.66 & 1.04e-02 & 2.01 & 4.67 & 0.78 & 2.87 & 4.58 \\ 
 \hline
[28.00,34.00) & [0.00,0.10) & 5.65e-04 & 11.39 & 4.97 & 3.88e-04 & 13.69 & 5.79 & 0.55 & 17.81 & 4.58 \\ 
 \hline
[28.00,34.00) & [0.10,0.20) & 1.81e-03 & 5.12 & 4.74 & 8.67e-04 & 8.86 & 5.79 & 0.42 & 10.24 & 4.58 \\ 
 \hline
[28.00,34.00) & [0.20,0.30) & 2.50e-03 & 4.06 & 4.88 & 1.20e-03 & 5.64 & 4.74 & 0.50 & 6.95 & 4.58 \\ 
 \hline
[28.00,34.00) & [0.30,0.40) & 3.24e-03 & 3.45 & 4.92 & 1.62e-03 & 5.02 & 4.65 & 0.55 & 6.09 & 4.58 \\ 
 \hline
[28.00,34.00) & [0.40,0.50) & 3.36e-03 & 3.36 & 4.61 & 2.11e-03 & 3.81 & 5.13 & 0.57 & 5.08 & 4.58 \\ 
 \hline
[28.00,34.00) & [0.50,2.00) & 1.34e-02 & 2.13 & 4.65 & 8.83e-03 & 2.00 & 4.61 & 0.65 & 2.93 & 4.58 \\ 
 \hline
[34.00,40.00) & [0.00,0.10) & 4.64e-04 & 8.56 & 5.41 & 2.63e-04 & 11.23 & 4.60 & 0.64 & 14.12 & 4.58 \\ 
 \hline
[34.00,40.00) & [0.10,0.20) & 1.18e-03 & 5.38 & 5.42 & 5.55e-04 & 7.33 & 4.84 & 0.51 & 9.09 & 4.58 \\ 
 \hline
[34.00,40.00) & [0.20,0.30) & 1.62e-03 & 4.20 & 4.74 & 7.23e-04 & 6.14 & 5.10 & 0.39 & 7.44 & 4.58 \\ 
 \hline
[34.00,40.00) & [0.30,0.40) & 1.83e-03 & 3.51 & 4.73 & 8.38e-04 & 5.63 & 4.74 & 0.47 & 6.64 & 4.58 \\ 
 \hline
[34.00,40.00) & [0.40,0.50) & 2.05e-03 & 3.44 & 4.68 & 1.09e-03 & 4.98 & 4.78 & 0.54 & 6.05 & 4.58 \\ 
 \hline
[34.00,40.00) & [0.50,2.00) & 8.81e-03 & 1.88 & 4.73 & 5.25e-03 & 2.36 & 4.69 & 0.59 & 3.02 & 4.58 \\ 
 \hline
[40.00,48.00) & [0.00,0.10) & 3.98e-04 & 9.82 & 4.70 & 1.38e-04 & 16.11 & 5.20 & 0.32 & 18.87 & 4.58 \\ 
 \hline
[40.00,48.00) & [0.10,0.20) & 7.99e-04 & 6.32 & 5.22 & 2.92e-04 & 10.10 & 4.90 & 0.44 & 11.92 & 4.58 \\ 
 \hline
[40.00,48.00) & [0.20,0.30) & 1.15e-03 & 4.63 & 4.79 & 4.77e-04 & 7.49 & 4.61 & 0.40 & 8.80 & 4.58 \\ 
 \hline
[40.00,48.00) & [0.30,0.40) & 1.33e-03 & 4.11 & 4.69 & 5.04e-04 & 7.10 & 5.62 & 0.44 & 8.21 & 4.58 \\ 
 \hline
[40.00,48.00) & [0.40,0.50) & 1.54e-03 & 3.83 & 4.93 & 6.55e-04 & 6.21 & 5.07 & 0.42 & 7.29 & 4.58 \\ 
 \hline
[40.00,48.00) & [0.50,2.00) & 7.02e-03 & 2.01 & 4.83 & 3.58e-03 & 2.76 & 4.63 & 0.50 & 3.41 & 4.58 \\ 
 \hline
[48.00,56.00) & [0.00,0.10) & 1.87e-04 & 11.91 & 4.80 & & & & & & \\ 
 \hline
[48.00,56.00) & [0.10,0.20) & 4.48e-04 & 10.33 & 4.99 & 1.55e-04 & 13.69 & 4.89 & 0.33 & 17.15 & 4.58 \\ 
 \hline
[48.00,56.00) & [0.20,0.30) & 7.06e-04 & 5.87 & 4.96 & 2.15e-04 & 10.82 & 5.22 & 0.37 & 12.31 & 4.58 \\ 
 \hline
[48.00,56.00) & [0.30,0.40) & 7.71e-04 & 5.31 & 4.92 & 2.19e-04 & 10.36 & 5.18 & 0.29 & 11.64 & 4.58 \\ 
 \hline
[48.00,56.00) & [0.40,0.50) & 7.46e-04 & 5.21 & 4.69 & 3.05e-04 & 8.80 & 4.93 & 0.39 & 10.23 & 4.58 \\ 
 \hline
[48.00,56.00) & [0.50,2.00) & 3.88e-03 & 2.46 & 4.67 & 1.80e-03 & 3.62 & 4.66 & 0.48 & 4.38 & 4.58 \\ 
 \hline
[56.00,68.00) & [0.00,0.10) & 1.90e-04 & 11.86 & 4.92 & & & & & & \\ 
 \hline
[56.00,68.00) & [0.10,0.20) & 4.24e-04 & 9.19 & 4.65 & 8.79e-05 & 16.67 & 5.70 & 0.19 & 19.03 & 4.58 \\ 
 \hline
[56.00,68.00) & [0.20,0.30) & 5.80e-04 & 6.28 & 5.96 & 1.71e-04 & 11.59 & 4.69 & 0.36 & 13.18 & 4.58 \\ 
 \hline
[56.00,68.00) & [0.30,0.40) & 6.27e-04 & 5.64 & 4.68 & 1.52e-04 & 11.85 & 5.76 & 0.30 & 13.12 & 4.58 \\ 
 \hline
[56.00,68.00) & [0.40,0.50) & 4.84e-04 & 6.22 & 4.76 & 1.57e-04 & 11.15 & 5.32 & 0.37 & 12.77 & 4.58 \\ 
 \hline
[56.00,68.00) & [0.50,2.00) & 2.81e-03 & 2.77 & 4.79 & 1.03e-03 & 4.51 & 5.04 & 0.36 & 5.29 & 4.58 \\ 
 \hline
[68.00,80.00) & [0.20,0.30) & 2.36e-04 & 10.32 & 5.45 & & & & & & \\ 
 \hline
[68.00,80.00) & [0.30,0.40) & 1.88e-04 & 10.39 & 4.89 & & & & & & \\ 
 \hline
[68.00,80.00) & [0.40,0.50) & 1.88e-04 & 10.15 & 4.80 & & & & & & \\ 
 \hline
[68.00,80.00) & [0.50,2.00) & 9.29e-04 & 4.69 & 4.75 & 3.33e-04 & 7.34 & 5.06 & 0.38 & 8.71 & 4.58 \\ 
 \hline

\end{longtable*}

\end{document}